\documentclass{aastex631}
\usepackage{amsmath}
\usepackage{placeins}
\newcommand\dfdt{${df}/{dt}$ }

\newcommand\Hzs{Hz~s$^{-1}$ }
\newcommand\Hzsns{Hz~s$^{-1}$}

\shorttitle{A Search for Technosignatures Around TESS exoplanets with the GBT}
\shortauthors{Margot et al.}

\graphicspath{{./}{figures/}}

\hypersetup{colorlinks=true,citecolor=blue,linkcolor=blue,urlcolor=blue,breaklinks=true}

\begin{document}
\title{A Search for Technosignatures Around 11,680 Stars with the Green Bank Telescope at 1.15--1.73 GHz}

\correspondingauthor{Jean-Luc Margot}
\email{jlm@epss.ucla.edu}

\author[0000-0001-9798-1797]{Jean-Luc Margot}
\affiliation{Department of Earth, Planetary, and Space Sciences, University of California, Los Angeles, CA 90095, USA}
\affiliation{Department of Physics and Astronomy, University of California, Los Angeles, CA 90095, USA}

\author[0000-0002-3012-4261]{Megan G. Li}
\affiliation{Department of Earth, Planetary, and Space Sciences, University of California, Los Angeles, CA 90095, USA}

\author[0000-0003-4736-4728]{Pavlo Pinchuk} 
\affiliation{National Renewable Energy Laboratory, Golden, CO 80401, USA}

\author[0000-0003-3994-5143]{Nathan Myhrvold}
\affiliation{Intellectual Ventures, 3150 139th Ave SE, Bellevue, WA 98005, USA}

\author[0000-0003-3970-688X]{Larry Lesyna}
\affiliation{LXL Technology, Las Vegas, NV 89148, USA}
\enlargethispage{0.5cm}

\enlargethispage{1cm}
\author{Lea E.  Alcantara}
\affiliation{UCLA Samueli School of Engineering, University of California, Los Angeles, CA 90095, USA}
\author{Megan T.  Andrakin}
\affiliation{Department of Physics and Astronomy, University of California, Los Angeles, CA 90095, USA}
\author{Jeth   Arunseangroj}
\affiliation{Department of Physics and Astronomy, University of California, Los Angeles, CA 90095, USA}
\author{Damien S.  Baclet}
\affiliation{Department of Mathematics, University of California, Los Angeles, CA 90095, USA}
\author{Madison H.  Belk}
\affiliation{Department of Electrical Engineering, University of California, Los Angeles, CA 90095, USA}
\author{Zerxes R.  Bhadha}
\affiliation{Department of Public Affairs, University of California, Los Angeles, CA 90095, USA}
\author{Nicholas W.  Brandis}
\affiliation{Department of Electrical and Computer Engineering, University of California, Los Angeles, CA 90095, USA}
\author{Robert E.  Carey}
\affiliation{Department of Electrical Engineering, University of California, Los Angeles, CA 90095, USA}
\author{Harrison P.  Cassar}
\affiliation{Department of Computer Science, University of California, Los Angeles, CA 90095, USA}
\author{Sai S.  Chava}
\affiliation{Department of Physics and Astronomy, University of California, Los Angeles, CA 90095, USA}
\author{Calvin   Chen}
\affiliation{Department of Mathematics, University of California, Los Angeles, CA 90095, USA}
\author{James   Chen}
\affiliation{Department of Electrical Engineering, University of California, Los Angeles, CA 90095, USA}
\author{Kellen T.  Cheng}
\affiliation{Department of Electrical Engineering, University of California, Los Angeles, CA 90095, USA}
\author{Alessia   Cimbri}
\affiliation{Department of Physics and Astronomy, University of California, Los Angeles, CA 90095, USA}
\author{Benjamin   Cloutier}
\affiliation{Department of Mathematics, University of California, Los Angeles, CA 90095, USA}
\author{Jordan A.  Combitsis}
\affiliation{Department of Electrical and Computer Engineering, University of California, Los Angeles, CA 90095, USA}
\author{Kelly L.  Couvrette}
\affiliation{Department of Materials Engineering, University of California, Los Angeles, CA 90095, USA}
\author{Brandon P.  Coy}
\affiliation{Department of Earth, Planetary, and Space Sciences, University of California, Los Angeles, CA 90095, USA}
\author{Kyle W.  Davis}
\affiliation{Department of Physics and Astronomy, University of California, Los Angeles, CA 90095, USA}
\author{Antoine F.  Delcayre}
\affiliation{Department of Physics and Astronomy, University of California, Los Angeles, CA 90095, USA}
\author{Michelle R.  Du}
\affiliation{Department of Electrical Engineering, University of California, Los Angeles, CA 90095, USA}
\author{Sarah E.  Feil}
\affiliation{Department of Earth, Planetary, and Space Sciences, University of California, Los Angeles, CA 90095, USA}
\author{Danning   Fu}
\affiliation{Department of Physics and Astronomy, University of California, Los Angeles, CA 90095, USA}
\author{Travis J.  Gilmore}
\affiliation{Department of Earth, Planetary, and Space Sciences, University of California, Los Angeles, CA 90095, USA}
\author{Emery   Grahill-Bland}
\affiliation{UCLA College of Letters and Science, University of California, Los Angeles, CA 90095, USA}
\author{Laura M.  Iglesias}
\affiliation{Department of Earth, Planetary, and Space Sciences, University of California, Los Angeles, CA 90095, USA}
\author{Zoe   Juneau}
\affiliation{Department of Physics and Astronomy, University of California, Los Angeles, CA 90095, USA}
\author{Anthony G.  Karapetian}
\affiliation{Department of Computer Science, University of California, Los Angeles, CA 90095, USA}
\author{George   Karfakis}
\affiliation{Department of Electrical Engineering, University of California, Los Angeles, CA 90095, USA}
\author{Christopher T.  Lambert}
\affiliation{Department of Earth, Planetary, and Space Sciences, University of California, Los Angeles, CA 90095, USA}
\author{Eric A.  Lazbin}
\affiliation{Department of Electrical Engineering, University of California, Los Angeles, CA 90095, USA}
\author{Jian H.  Li}
\affiliation{Department of Mechanical Engineering, University of California, Los Angeles, CA 90095, USA}
\author{Zhuofu (Chester)  Li}
\affiliation{Department of Physics and Astronomy, University of California, Los Angeles, CA 90095, USA}
\author{Nicholas M.  Liskij} %
\affiliation{Department of Mathematics, University of California, Los Angeles, CA 90095, USA}
\author{Anthony V.  Lopilato} %
\affiliation{Department of Electrical Engineering, University of California, Los Angeles, CA 90095, USA}
\author{Darren J.  Lu}
\affiliation{Department of Computer Science, University of California, Los Angeles, CA 90095, USA}
\author{Detao   Ma}
\affiliation{Department of Electrical Engineering, University of California, Los Angeles, CA 90095, USA}
\author{Vedant   Mathur}
\affiliation{Department of Electrical and Computer Engineering, University of California, Los Angeles, CA 90095, USA}
\author{Mary H.  Minasyan}
\affiliation{Department of Physics and Astronomy, University of California, Los Angeles, CA 90095, USA}
\author{Maxwell K.  Muller} %
\affiliation{Department of Physics and Astronomy, University of California, Los Angeles, CA 90095, USA}
\author{Mark T.  Nasielski}
\affiliation{Department of Electrical and Computer Engineering, University of California, Los Angeles, CA 90095, USA}
\author{Janice T.  Nguyen}
\affiliation{Department of Physics and Astronomy, University of California, Los Angeles, CA 90095, USA}
\author{Lorraine M.  Nicholson}
\affiliation{Department of Physics and Astronomy, University of California, Los Angeles, CA 90095, USA}
\author{Samantha   Niemoeller} %
\affiliation{Department of Computer Science, University of California, Los Angeles, CA 90095, USA}
\author{Divij   Ohri}
\affiliation{Department of Computer Science, University of California, Los Angeles, CA 90095, USA}
\author{Atharva U.  Padhye}
\affiliation{Department of Electrical Engineering, University of California, Los Angeles, CA 90095, USA}
\author{Supreethi V.  Penmetcha}
\affiliation{Department of Mechanical Engineering, University of California, Los Angeles, CA 90095, USA}
\author{Yugantar   Prakash}
\affiliation{Department of Mathematics, University of California, Los Angeles, CA 90095, USA}
\author{Xinyi (Cindy) Qi}
\affiliation{Department of Physics and Astronomy, University of California, Los Angeles, CA 90095, USA}
\author{Liam Rindt}  %
\affiliation{Department of Physics and Astronomy, University of California, Los Angeles, CA 90095, USA}
\author{Vedant   Sahu}
\affiliation{Department of Physics and Astronomy, University of California, Los Angeles, CA 90095, USA}
\author{Joshua A.  Scally}
\affiliation{Department of Physics and Astronomy, University of California, Los Angeles, CA 90095, USA}
\author{Zefyr   Scott}
\affiliation{Department of Electrical Engineering, University of California, Los Angeles, CA 90095, USA}
\author{Trevor J.  Seddon}
\affiliation{Department of Physics and Astronomy, University of California, Los Angeles, CA 90095, USA}
\author{Lara-Lynn V.  Shohet}
\affiliation{Department of Physics and Astronomy, University of California, Los Angeles, CA 90095, USA}
\author{Anchal   Sinha}
\affiliation{Department of Electrical and Computer Engineering, University of California, Los Angeles, CA 90095, USA}
\author{Anthony E.  Sinigiani}
\affiliation{Department of Aerospace Engineering, University of California, Los Angeles, CA 90095, USA}
\author{Jiuxu   Song}
\affiliation{Department of Electrical and Computer Engineering, University of California, Los Angeles, CA 90095, USA}
\author{Spencer M.  Stice}
\affiliation{Department of Electrical and Computer Engineering, University of California, Los Angeles, CA 90095, USA}
\author{Nadine M.  Tabucol} %
\affiliation{Department of Physics and Astronomy, University of California, Los Angeles, CA 90095, USA}
\author{Andria   Uplisashvili}
\affiliation{Department of Physics and Astronomy, University of California, Los Angeles, CA 90095, USA}
\author{Krishna   Vanga}
\affiliation{Department of Electrical Engineering, University of California, Los Angeles, CA 90095, USA}
\author{Amaury G.  Vazquez}
\affiliation{Department of Physics and Astronomy, University of California, Los Angeles, CA 90095, USA}
\author{George   Vetushko}
\affiliation{Department of Neuroscience, University of California, Los Angeles, CA 90095, USA}
\author{Valeria   Villa}
\affiliation{Department of Earth, Planetary, and Space Sciences, University of California, Los Angeles, CA 90095, USA}
\author{Maria   Vincent}
\affiliation{Department of Earth, Planetary, and Space Sciences, University of California, Los Angeles, CA 90095, USA}
\author{Ian J.  Waasdorp}
\affiliation{Department of Physics and Astronomy, University of California, Los Angeles, CA 90095, USA}
\author{Ian B.  Wagaman}
\affiliation{Department of Physics and Astronomy, University of California, Los Angeles, CA 90095, USA}
\author{Amanda   Wang}
\affiliation{Department of Computer Science, University of California, Los Angeles, CA 90095, USA}
\author{Jade C.  Wight}
\affiliation{Department of Earth, Planetary, and Space Sciences, University of California, Los Angeles, CA 90095, USA}
\author{Ella   Wong}
\affiliation{Department of Physics and Astronomy, University of California, Los Angeles, CA 90095, USA}
\author{Natsuko   Yamaguchi}
\affiliation{Department of Physics and Astronomy, University of California, Los Angeles, CA 90095, USA}
\author{Zijin   Zhang}
\affiliation{Department of Earth, Planetary, and Space Sciences, University of California, Los Angeles, CA 90095, USA}
\author{Junyang   Zhao}
\affiliation{Department of Electrical Engineering, University of California, Los Angeles, CA 90095, USA}
\author[0000-0001-5229-7430]{Ryan S.\ Lynch}
\affiliation{Green Bank Observatory, P.O.\ Box 2, Green Bank, WV 24494, USA}

\begin{abstract}
We conducted a search for narrowband radio signals over four observing
sessions in 2020--2023 with the L-band receiver (1.15--1.73 GHz) of
the 100 m diameter Green Bank Telescope.  We pointed the telescope in
the directions of 62 TESS Objects of Interest, capturing radio
emissions from a total of $\sim$11,680 stars and planetary systems in
the $\sim$9 arcminute beam of the telescope.  All detections were
either automatically rejected or visually inspected and confirmed to
be of anthropogenic nature.  We also quantified the end-to-end
efficiency of radio SETI pipelines with a signal injection and
recovery analysis.  The UCLA SETI pipeline recovers $94.0\%$ of the
injected signals over the usable frequency range of the receiver and
$98.7\%$ of the injections when regions of dense radio frequency
interference are excluded.  In another pipeline that uses incoherent
sums of 51 consecutive spectra, the recovery rate is $\sim$15 times
smaller at $\sim$6\%.  The pipeline efficiency affects calculations of
transmitter prevalence and SETI search volume.  Accordingly, we
developed an improved Drake Figure of Merit and a formalism to place
upper limits on transmitter prevalence that take the pipeline
efficiency and transmitter duty cycle into account.  Based on our
observations, we can state at the 95\% confidence level that fewer
than 6.6\% of stars within 100~pc host a transmitter that is
detectable in our search (EIRP $>$ 10$^{13}$~W).  For stars within
20,000~ly, the fraction of stars with detectable transmitters (EIRP
$>$ 5 $\times$ 10$^{16}$~W) is at most $3 \times 10^{-4}$.  Finally,
we showed that the UCLA SETI pipeline natively detects the signals
detected with AI techniques by \citet{ma23}.
\end{abstract}

\keywords{Search for extraterrestrial intelligence - technosignatures - astrobiology — exoplanets - radio astronomy - Milky Way Galaxy}

\vfill
\section{Introduction}
\label{sec-intro}

In the 1982 decadal report, the Astronomy Survey Committee of the National Research
Council (NRC) recommended the approval and funding of ``An
astronomical Search for Extraterrestrial Intelligence (SETI),
supported at a modest level, undertaken as a long-term effort rather
than as a short-term project, and open to the participation of the
general scientific community'' \citep{nas82}.
The Committee noted:
\begin{quote}It is hard to imagine a more exciting astronomical discovery or one that
would have greater impact on human perceptions than the detection of
extraterrestrial intelligence. After reviewing the arguments for and against SETI,
the Committee has concluded that the time is ripe for initiating a modest program
that might include a survey in the microwave region of the electromagnetic
spectrum while maintaining an openness to support of other innovative studies as
they are proposed.
\end{quote}

In a subsequent report on the search for life's origins, the NRC
stated: ``Two parallel avenues of research should be pursued in
attempts to detect life beyond the solar system: searches for evidence
of biological modification of an extrasolar planet and searches for
evidence of extraterrestrial technology'' \citep{nas90life}.  The
report's recommendations included the ``commencement of a systematic
ground-based search through the low end of the microwave window for
evidence of signals from an extraterrestrial technology''.

The detection of extraterrestrial life forms
is expected to usher  profound
developments in a wide range of scientific and cultural disciplines.
These potential benefits provide compelling incentives to invest in
multifaceted searches for biological indicators  ({\em
  biosignatures}) and technological indicators ({\em technosignatures}) of 
extraterrestrial life.
Searches for biosignatures and technosignatures are highly
complementary.  In particular, the latter can ``expand the search for
life in the universe from primitive to complex life and from the solar
neighborhood to the entire Galaxy'' \citep{marg19astro2020}.  In the
Milky Way Galaxy alone, the ratio of search volumes with current and
near-future technology is $V_{\rm techno} / V_{\rm bio} \gtrsim 10^6$.
In terms of the number of accessible targets, the ratio is $N_{\rm
  techno} / N_{\rm bio} \gtrsim 10^9$.

Although some types of solar system biosignatures (e.g., a fossil or
sample organism) may offer compelling interpretations, the proposed
exoplanet biosignatures are expected to
yield inconclusive interpretations for some time
\citep[e.g.,][]{fuji18others}.  Abiogenic interpretations
may remain difficult to rule out~\citep[e.g.,][]{rein14}, as evidenced
by biosignature claims for planets that are a million times closer
than the nearest exoplanet (methane on Mars, phosphine on Venus).  In
many cases, the spectroscopic observations may be consistent with but
not diagnostic of the presence of
life~\citep[e.g.,][]{catl18others,mead22others}.  In contrast, the
search for technosignatures provides an opportunity to obtain robust
detections with unambiguous interpretations.  An example of such a
technosignature is a narrowband (say, $<$10~Hz at gigahertz
frequencies) signal from an emitter located beyond the solar system.
Detection of a signal with these characteristics would provide
sufficient evidence for the existence of another civilization because
natural settings cannot generate such signals.  In order to confine
the signal bandwidth within 10 Hz at L band, the velocity dispersion
and Doppler broadening of the species participating in the emission
must remain below 2 m/s.
Such coherence in velocity would have to be
maintained over the physical scales of the emission sites.
Fluid astrophysical settings cannot produce such conditions because
the thermal velocity of species is much larger, even in the coldest
environments.  Even astrophysical masers cannot maintain this degree
of coherence: the narrowest reported OH (1612 MHz) maser
line width is 550 Hz~\citep{cohe87others, qiao20others}, roughly two
orders of magnitude wider than the proposed narrowband radio
technosignatures.

Here, we describe a search for narrowband radio technosignatures
around $\sim$11,680 stars and their planetary systems.  For the
historical context of the search, see, e.g., \citet{Tarter2001,Tarter2010,technosigs18}.

\section{Observations}
\label{sec-sources}
We observed $\sim$11,680 stars and their planetary systems in 62
distinct directions aligned with TESS Objects of Interest (TOIs).
The characteristics of our primary targets are listed in
Appendix~\ref{app-sources} (Tables~\ref{tab-targets1} and \ref{tab-targets2}).  To compute the number of stars captured
by the 8.4 arcmin beamwidth of the telescope at 1.42 GHz, we followed
\citet{wlod20} and performed cone searches with the Gaia catalog
\citep{gaiadr3}.  We found 11,680 known stars, of which
10,378 have improved geometric distance measurements calculated by
\citet{bail21}.  The median, mean, and maximum distance estimates for
these sources are 2288 pc (7461 ly), 2450 pc (7990 ly), and 12,664 pc
(41,305 ly), respectively.  There are 10,230 sources located within 6132 pc
(20,000 ly) of the Sun.

We observed these stars and their planetary systems with the Green
Bank Telescope (GBT) during 2 hr sessions on 2020 April 22, 2021
April 28, 2022 May 22, and 2023 May 13.  The observing cadence
consisted of two scans of 150~s each per primary target, with sources
arranged in pairs resulting in an A-B-A-B sequence for sources A and
B.  Angular separations between sources always exceeded several
telescope beamwidths.  These ON-OFF-ON-OFF (or OFF-ON-OFF-ON)
sequences are particularly useful in the detection of radio frequency
interference (RFI) (Section~\ref{sec-filters}).

We recorded both linear polarizations of the L-band
receiver with the VEGAS backend in its baseband recording mode
\citep{VEGAS}.  We sampled 800 MHz of bandwidth between 1.1 and
1.9~GHz.
We sampled complex (in-phase
and quadrature) voltages with 8-bit quantization, but preserved only
2-bit samples after requantization with an optimal four-level sampler,
which yields a quantization efficiency $\eta_Q$ of 0.8825
\citep{Kogan1998}.

\section{Methods}
\label{sec-methods}

Our data processing techniques are generally similar to those used by \citet{marg18seti}, \citet{Pinchuk2019}, and \citet{marg21setiothers}.  Here, we give a brief overview and refer the reader to these other works for additional details.

\subsection{Bandpass Correction}
\label{sec-scale}
The VEGAS instrument splits the 800 MHz recorded bandwidth into 256
coarse channels of 3.125 MHz each.  In the process of doing so, it applies a
bandpass filter to each coarse channel.  This filter reduces the
amplitude of the baseline near both edges of the spectra.  We restored
an approximately flat baseline by dividing each spectrum by a model of
the bandpass filter response.  This model was obtained by fitting a
16-degree Chebyshev polynomial to the median bandpass response of 28
scans in the 1664.0625--1667.1875 MHz frequency range, which is generally devoid
of interference because it falls in the middle of the radio astronomy
protected band (1660.6--1670.0 MHz) for the hydroxyl radical.  We
enforced an even response by setting the odd coefficients of the
polynomial to zero.

\subsection{Doppler Dechirping}
\label{sec-tree}

Over the 150 s duration of our scans, narrowband signals from
fixed-frequency transmitters are well approximated at the receiver by
linear frequency modulated (FM) ``chirp'' signals, where the rate of
change in frequency is dictated by the orbital and rotational motions
of both the emitter and the receiver.
The linear FM waveform is characterized by a signal of the form
\begin{equation}
s(t) = A \cos(2\pi (f_0 t + K t^2/2)),  \hspace{2cm} 0 \leq t \leq \tau,
\end{equation}
where $A$ is the signal amplitude, $f_0$ is the frequency at $t=0$,
$K$ is the rate of change of the frequency, and $\tau$ is the duration of
the scan.  In complex exponential notation,
\begin{equation}
  s(t) = A \exp{(j2\pi (f_0 t + K t^2/2))} = A \exp{(j \theta(t))},  \hspace{2cm} 0 \leq t \leq \tau.
  \label{eq-chirp}
\end{equation}
The instantaneous frequency is the time derivative of the phase, i.e.,
\begin{equation}
  f(t) = \frac{1}{2\pi} \frac{d\theta(t)}{dt} = f_0 + K t. 
\end{equation}
The frequency increases linearly as $f(t) = f_0 + Kt$, with a total
frequency excursion equal to $K \tau$.  An
audible signal with this time-frequency behavior would sound like a
chirp, hence the name commonly attributed to the waveform.

Doppler dechirping consists of compensating for a signal's drift in
time-frequency space to facilitate integration of the signal power
over the scan duration.  In SETI searches, the drift rate is not known
a priori.  We used a tree algorithm of complexity O($N \log N$)
\citep{Taylor1974, Siemion2013} to integrate the signal power at 1023
trial drift rates over the range $\pm8.86$ \Hzs with a drift rate
resolution of $\Delta \dot{f} = 0.0173$~\Hzsns.  This approximate
technique, known as {\em incoherent dechirping}, does not recover
100\% of the signal power.  \citet{marg21setiothers} quantified this
signal loss with a dechirping efficiency factor $\eta_D$ $(0 \leq
\eta_D \leq 1)$ for a variety of settings, including searches that
utilize incoherent summing of power spectra prior to signal detection.
In this and previous UCLA SETI work, we do not use incoherent
averaging (i.e., $N_{\rm INC\_SUMS}=1$) and the dechirping efficiency
ranges between 60\% and 100\% with an average $\eta_D \simeq 72\%$
over the $\pm8.86$ \Hzs drift rate range.  For searches with $N_{\rm
  INC\_SUMS}=51$ over a $\pm4$ \Hzs drift rate range \citep[e.g.,][]{pric20,gajj21} the dechirping
efficiency ranges between 4\% and 100\% with an average $\eta_D \simeq
16\%$.  Importantly, the nominal performance of the tree algorithm for
such searches is maintained in only a fairly narrow range of drift
rates up to $\pm0.15$ \Hzsns, and the efficiency drops precipitously
at larger drift rates due to Doppler smearing of the signal
\citep[][Figure 7]{marg21setiothers}.

\begin{figure}[h!]
\begin{center}
  \includegraphics[width=0.6\textwidth]{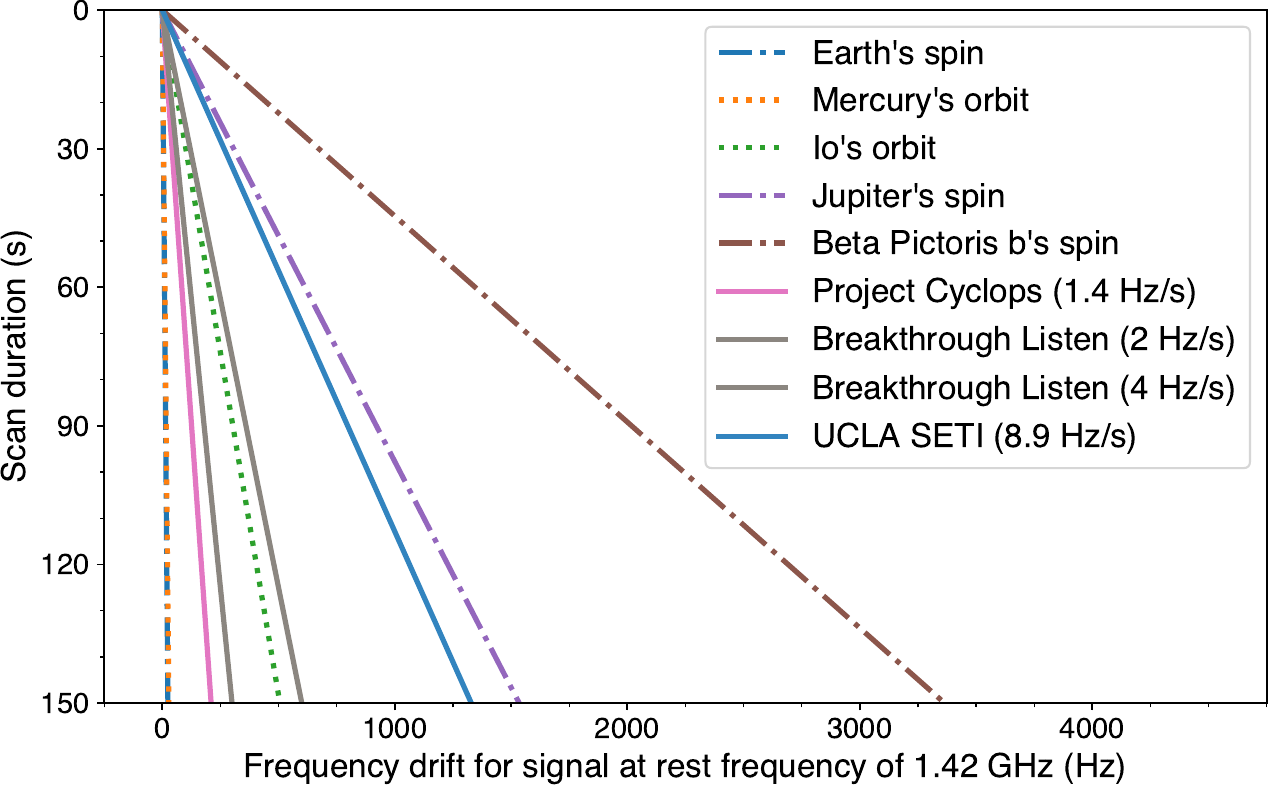}
\end{center}
\caption{Expected frequency drifts from monochromatic ($f_t$ = 1.42 GHz)
  transmitters
  experiencing spin and orbital accelerations in various settings.
  The solid lines represent the maximum drift rates corresponding to
  Project Cyclops, Breakthrough Listen (BL), and UCLA SETI searches.  This figure is adapted
  from \citet{Sheikh2019} and reduces the drift rates sampled by BL by
  a factor of $\sim$7 to correct an error in the original figure.}
  \label{fig-drifts}
\end{figure}

The received frequency
$f_r$
of a monochromatic transmission  at
frequency
$f_t$
experiences a time rate of change that
depends on the line-of-sight acceleration
$\dot{v}$
between transmitter and
  receiver.  To first order,
\begin{equation}
  \frac{\dot{f}_r}{f_t} = \frac{\dot{v}}{c},
  \label{eq-dfdt}
\end{equation}
where the overdot denotes a time derivative and $c$ is the speed of light.
Our selection of a range of trial drift rates with maximum value $\dot{f}_{r,\rm max} = \pm8.86$ \Hzs 
corresponds to
a fractional drift rate of $\pm$6.24 nHz at $f_t$ = 1.42 GHz (maximum
accelerations $\dot{v}_{\rm max}$ of 1.87~ms$^{-2}$).  It is an
appropriate choice because it accommodates line-of-sight accelerations
due to the spins and orbits of most exoplanets.  It can handle
accelerations due the orbits of $\sim$73\% of confirmed exoplanets
with known semi-major axes and orbital periods, $\sim$93\% of
confirmed exoplanets with semi-major axes greater than 0.05 au, and
100\% of confirmed exoplanets with semi-major axes greater than 0.1
au \citep{cptable}.  It can also handle accelerations due to the spins of Earth-size
planets at arbitrary periods (above the rotational breakup period) and
Jupiter-size planets with spin periods greater than 11.5~hr.
Transmitters located on exotic platforms that somehow exceed these
limits
could escape detection by our pipeline if the transmitted waveforms
were not compensated to account for the platform's acceleration
(Figure~\ref{fig-drifts}).

\subsection{Signal Detection}
At each frequency bin, our algorithm selects the trial drift rate that
yields the greatest integrated signal power and determines whether the
{\em prominence} of the signal \citep{marg21setiothers} exceeds 10
times the standard deviation of the noise.  The properties of signals
that exceed the threshold are stored in a structured query language
(SQL) database.

In practice, the algorithm proceeds in order of
decreasing prominence in each coarse channel.

\subsection{Doppler and Direction-of-origin Filters}
\label{sec-filters}
Signals with
$\dot{f}_r=0$ are marked as anthropogenic RFI because they
imply zero line-of-sight acceleration between the transmitter and
receiver.  Signals that are detected in more than one direction on the
sky are also marked as RFI because a signal emitted beyond the solar
system can appear in only one telescope beam.  Finally, sources that
are detected in only one of the two scans are also marked as
intermittent RFI.

The direction-of-origin filter, also known as a directional filter or
sky localization filter, can be run efficiently by retrieving the
signal properties from our SQL database.  A more stringent filter can
be obtained by running the machine-learning (ML) algorithm of
\citet{pinc22}.

\subsection{Visual Inspection of the Remaining Signals}
Signals that remain after the line-of-sight distance and acceleration
elimination process are marked as candidate technosignatures.  All
such candidates that fall outside of permanent RFI bands \citep{Pinchuk2019} are
visually inspected.

\subsection{Sensitivity}

The flux from a transmitter with equivalent isotropic radiated power (EIRP) at distance $r$ is
$S = {\rm EIRP} / (4\pi r^2).$

The signal-to-noise ratio (S/N) of a narrowband radio link has been computed by, e.g., \citet{frii46,krau86,enri17,marg18seti}.  It reads 
\begin{equation}
S/N = \frac{S}{\rm SEFD} \sqrt{\frac{n_{\rm pol} \tau}{\Delta f}}, 
\label{eq-snr}
\end{equation}
where 
$S$ is the observed flux,
SEFD is the system equivalent flux density, a common measure of telescope and receiver performance,
$n_{\rm pol}$ is the number of polarizations summed incoherently, 
$\tau$ is the integration time, 
and $\Delta f$ is the channel receiver bandwidth (i.e., frequency resolution).

In a more rigorous formulation, the S/N includes the quantization
efficiency $\eta_Q$ due to imperfect digitization of the voltage signals and
the dechirping efficiency $\eta_D$ due to imperfect integration of the signal
power \citep{marg21setiothers}:
\begin{equation}
S/N = \eta_Q \eta_D  \frac{S}{\rm SEFD} \sqrt{\frac{n_{\rm pol} \tau}{\Delta f}}.
\label{eq-snrbetter}
\end{equation}
Quantization efficiency approaches unity with 8-bit sampling and is
$\eta_Q$ = 88.25\% for an optimal 2-bit sampler \citep{Kogan1998}.
Dechirping efficiency with an O($N \log N$) algorithm depends on the
frequency drift rate and ranges between 60\% and 100\% for the data
acquisition and processing choices in this and previous UCLA SETI
work, and between 4\% and 100\% for a BL-like process with $N_{\rm
  INC\_SUMS}=51$ and $\dot{f}_r$ within $\pm$4 \Hzsns.  It is
possible to improve the dechirping efficiency if one is willing to use
a costly O($N^2$) incoherent dechirping algorithm.  For signals of
interest with known frequency drift rates, UCLA SETI has the
capability to apply a coherent dechirping algorithm to the raw voltage
data, in which case $\eta_D \simeq 1$.

For the UCLA SETI program at the GBT,
we have $\eta_Q$ = 0.8825,
SEFD = 10 Jy,
$n_{\rm pol}$ = 2,
$\tau$ = 150 s,
and $\Delta f$ $\simeq$ 3~Hz.
Our usual detection threshold is set at S/N=10, such that signals with
flux $S_{\rm det} = 11.3 \times 10^{-26}$ W/m$^2$ are detectable.
With these parameters, an Arecibo Planetary Radar (EIRP=$2.2\times
10^{13}$ W) is detectable at 415 ly and a thousand Arecibos can be
detected at 13,123 ly.
Conversely, transmitters located
326 ly (100 pc) away are detectable with 0.62 Arecibos (EIRP=$1.35\times10^{13}$ W),
transmitters located 20,000 ly (6132 pc) away are detectable with 2323 Arecibos (EIRP=$5.08\times
10^{16}$ W), and 
transmitters located at the galactic
center are detectable with 4130 Arecibos.

\subsection{Signal Injection and Recovery Analysis}
\label{sec-injection-methods}

To quantify the end-to-end efficiency of the UCLA SETI pipeline, we
injected 10,000 artificial chirp signals in raw voltage data from our
2021 search, processed the data as we normally do, and quantified the
number of injected signals that were recovered by the pipeline.

\begin{figure}[ht!]
  \begin{tabular}{cc}
    \includegraphics[width=0.45\textwidth]{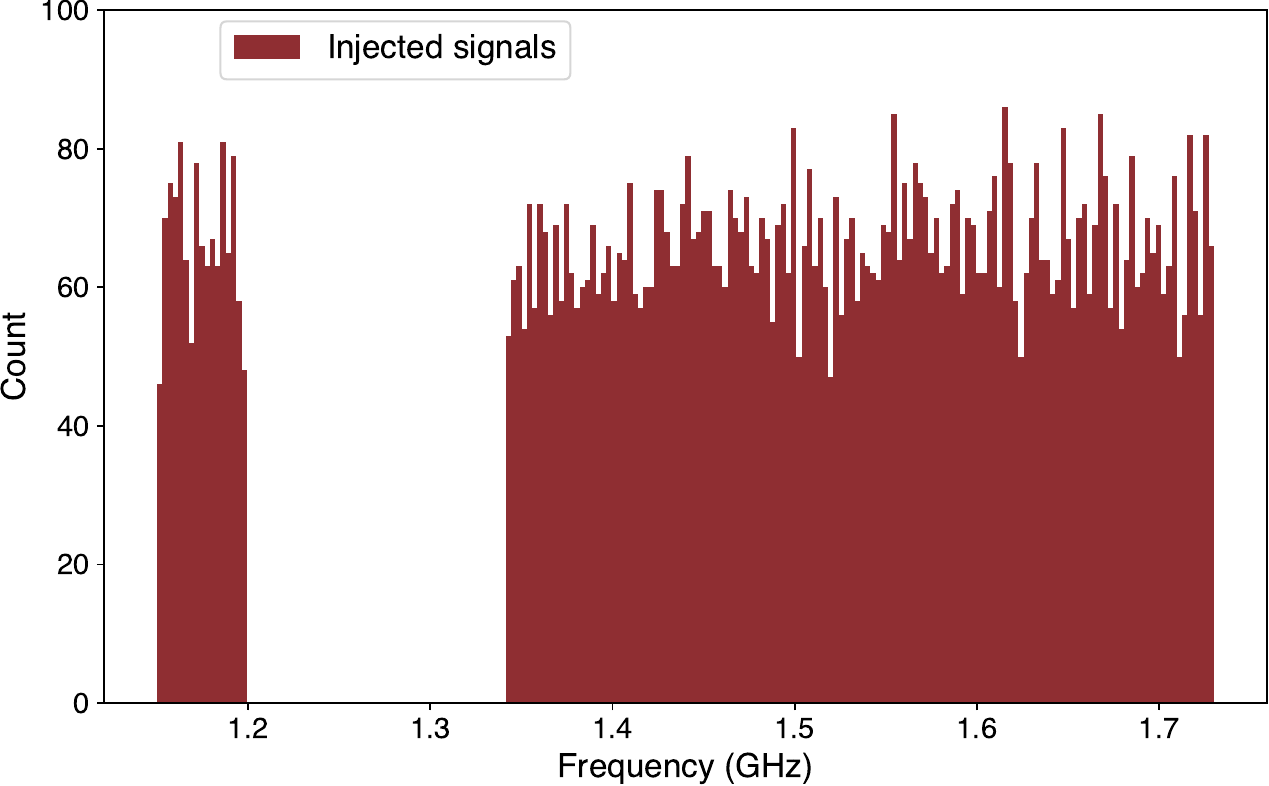} &
    \includegraphics[width=0.45\textwidth]{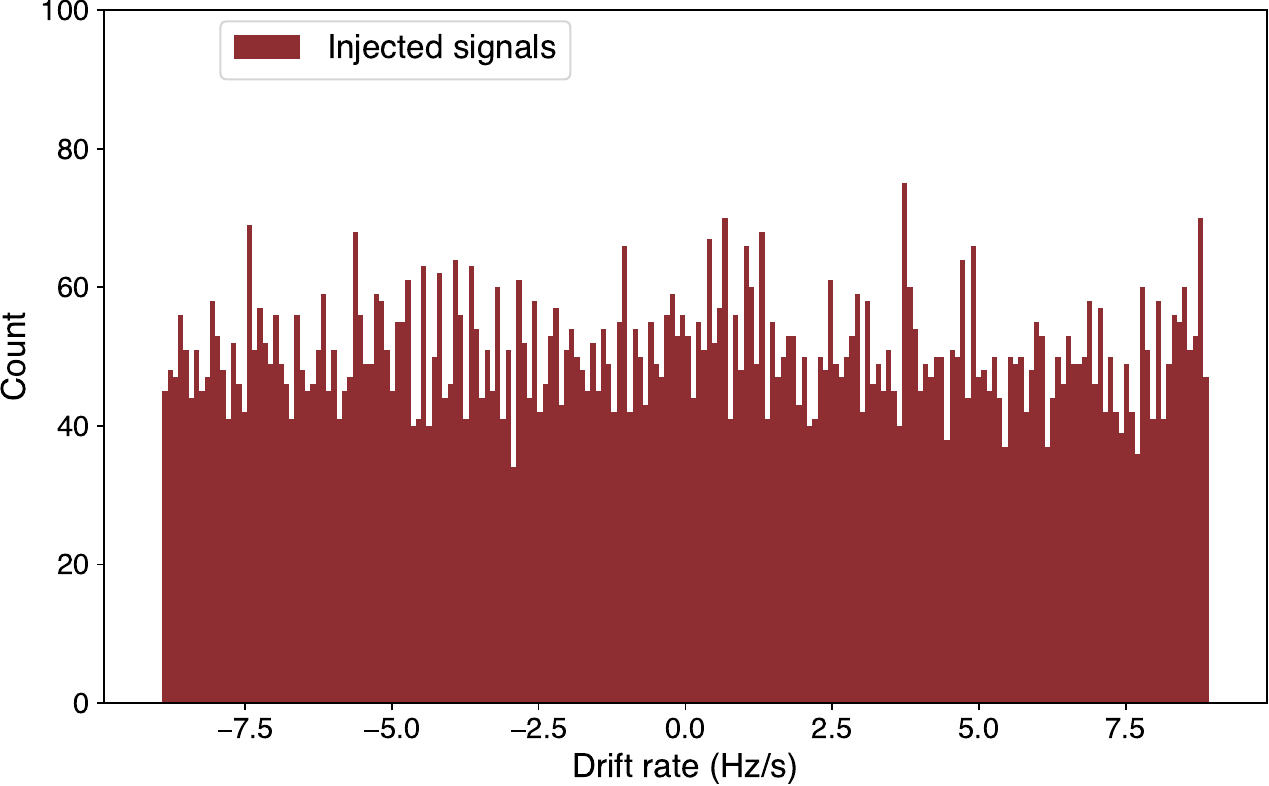}
  \end{tabular}
    \caption{Distributions of starting frequencies (Left) and drift rates
      (Right) of 10,000 artificial, linear FM ``chirp'' signals injected in voltage data.
      Frequencies corresponding to a notch filter at the GBT (1.2 -- 1.3412 GHz) are excluded.
    }
  \label{fig-inject}      
\end{figure}

We used Equation (\ref{eq-chirp}) to inject the artificial signals in
complex voltage data sampled with 8 bit quantization, and we adjusted
the signal amplitudes to achieve an S/N upon recovery of approximately
20.  The starting frequencies of the signals were drawn randomly from
a uniform distribution in the range 1.15--1.73 GHz, with the exclusion
of the 1.20--1.34 GHz range that is blocked by a notch filter at the
GBT.  The drift rates were drawn randomly from a uniform distribution
in the range $\pm8.86$ \Hzs (Figure~\ref{fig-inject}).

We used the exact same data files (raw voltage data files injected
with 10,000 artificial signals) to estimate the end-to-end efficiency
of a process that imitates the Breakthrough Listen (BL) pipeline.
Specifically, we computed power spectra with the FFTW library
\citep{fftw05} and a transform length of $2^{20}$, yielding a
frequency resolution $\Delta f = 2.98$~Hz that approximates the finest
frequency resolutions (2.79~Hz, 2.84~Hz, 2.93~Hz) of BL spectra
~\citep[][Table 4]{lebo19}.  We applied a bandpass correction
appropriate for the BL data acquisition backend.  We then summed 51
consecutive power spectra incoherently, yielding a time resolution of
17.11~s, to approximate the 51-fold incoherent summing and time
resolutions (17.40~s, 17.98~s, 18.25~s) of the High Spectral
Resolution (HSR) BL spectra~\citep[][Table 4]{lebo19}.  Finally, we
ran BL's version of the Doppler dechirping tree algorithm, as
implemented in turboSETI \citep{enri17}, with a maximum drift
rate of $\pm8.88$~\Hzs and minimum S/N of 10, to identify signals and
quantify the number of injected signals that were recovered ({\tt
  turboSETI -M 8.881784197 -s 10}).

A signal was deemed to be recovered if two conditions were jointly
met: (1) the recovered frequency was within $\pm$6 Hz of the injected
frequency, and (2) the recovered drift rate was within $\pm$0.05 \Hzs
of the injected drift rate.  These tolerances were designed to
accommodate slight mismatches of up to two bins in the frequency
dimension and the frequency drift rate dimension.  The probability of an
accidental match is less than one in a billion.  Because of differences
in algorithm implementation, the drift rate bins were 0.0173 \Hzs for
UCLA SETI and 0.0249 \Hzs for turboSETI.

\subsection{Native Detections of ML Candidates by the UCLA SETI Pipeline}
\citet{ma23} used a $\beta$-convolutional variational autoencoder and
random forest analysis to identify eight promising signals not previously
identified by the BL pipeline.  They named these candidates MLc1 to
MLc8, in reference to the machine learning (ML) process used in their
analysis.  We were interested in finding out whether these signals
could be natively detected by the UCLA SETI pipeline, without any ML
assistance.  We downloaded the BL HSR power spectra of the MLc
candidates and corresponding OFF scans, applied a bandpass correction
(Section \ref{sec-scale}) appropriate for the BL data acquisition
backend, and ran the resulting spectra through the UCLA SETI pipeline.
We used only the first four of six scans of each pair of sources to
mimic the A-B-A-B sequence used in UCLA SETI observations.

\section{Results}
\label{sec-results}

\subsection{Candidate Technosignatures}
We detected 41.2 million narrowband signals with the data from our
2020--2023 observations.  Almost all
(99.43\%)
of these signals were
rejected automatically by the UCLA SETI pipeline as RFI, which left
approximately 230,000 signals warranting further consideration.  Tens
of thousands of these signals are being inspected by thousands of
volunteers on the website
\href{http://arewealone.earth}{http://arewealone.earth} as part of a
citizen science collaboration \citep{li23aas}.  Almost all (99.78\%)
of the remaining signals were detected in regions of dense RFI.  We
visually inspected all $\sim$500 candidate technosignatures that were
detected outside of dense RFI regions and
determined that they were all anthropogenic.

It is remarkable that, in over 82 million narrowband signal detections
obtained during an 8-year period (Table \ref{tab-stats}), not a single
signal has merited follow-up observations.  There have been plenty of
instances where promising signals were only marginally detected in the
corresponding OFF scans.  We eliminate such signals from
consideration.

\begin{table}
  \begin{center}
\begin{tabular}{llrrrrrr}
  Data Set & Fields & Targets & Stars & Signals & Hit Rate Density  & DFM  & MDFM\\
  & & (primary) & (in beam) & (millions) & (sig kHz$^{-1}$ hr$^{-1}$) & (GHz m$^3$ W$^{-3/2}$) & (Hz$^2$ m$^3$ W$^{-3/2}$) \\
  \hline
  UCLA 2016 & Kepler  & 14       & 11,658 & 5.22 & 10.2 & 6.74e+31 & 3.95e+32\\
  UCLA 2017 & Kepler+ & 12       &  6,924 & 8.52 & 16.2 & 6.35e+31 & 3.72e+32\\
  UCLA 2018–19 & Gal. plane & 30 & 25,293 & 27.0 & 24.6 & 1.44e+32 & 8.47e+32\\
  UCLA 2020–23 & TESS & 62       & 11,680 & 41.2 & 18.2 & 2.99e+32 & 1.75e+33\\ %
\hline
  Total 2016–23 &     & 118      & 55,555 & 82.0 & 19.0 & 5.74e+32 & 3.37e+33\\
\end{tabular}
  \end{center}
\caption{UCLA SETI search characteristics, showing observation fields, number of primary targets, number of stars observed in the beam of the telescope, number of narrowband signals detected with S/N$>$10, hit rate density (number of detections per unit bandwidth per unit on-source time), Drake Figure of Merit (DFM), and modified DFM (Section~\ref{sec-fom}).  Properties of all 2016--2019 detections are available online \citep{seti16dataset,seti17dataset,seti18dataset}.}
\label{tab-stats}
\end{table}

\subsection{Signal Injection and Recovery Analysis}
\label{sec-ir}

\begin{figure}[h!]
\includegraphics[width=0.9\textwidth]{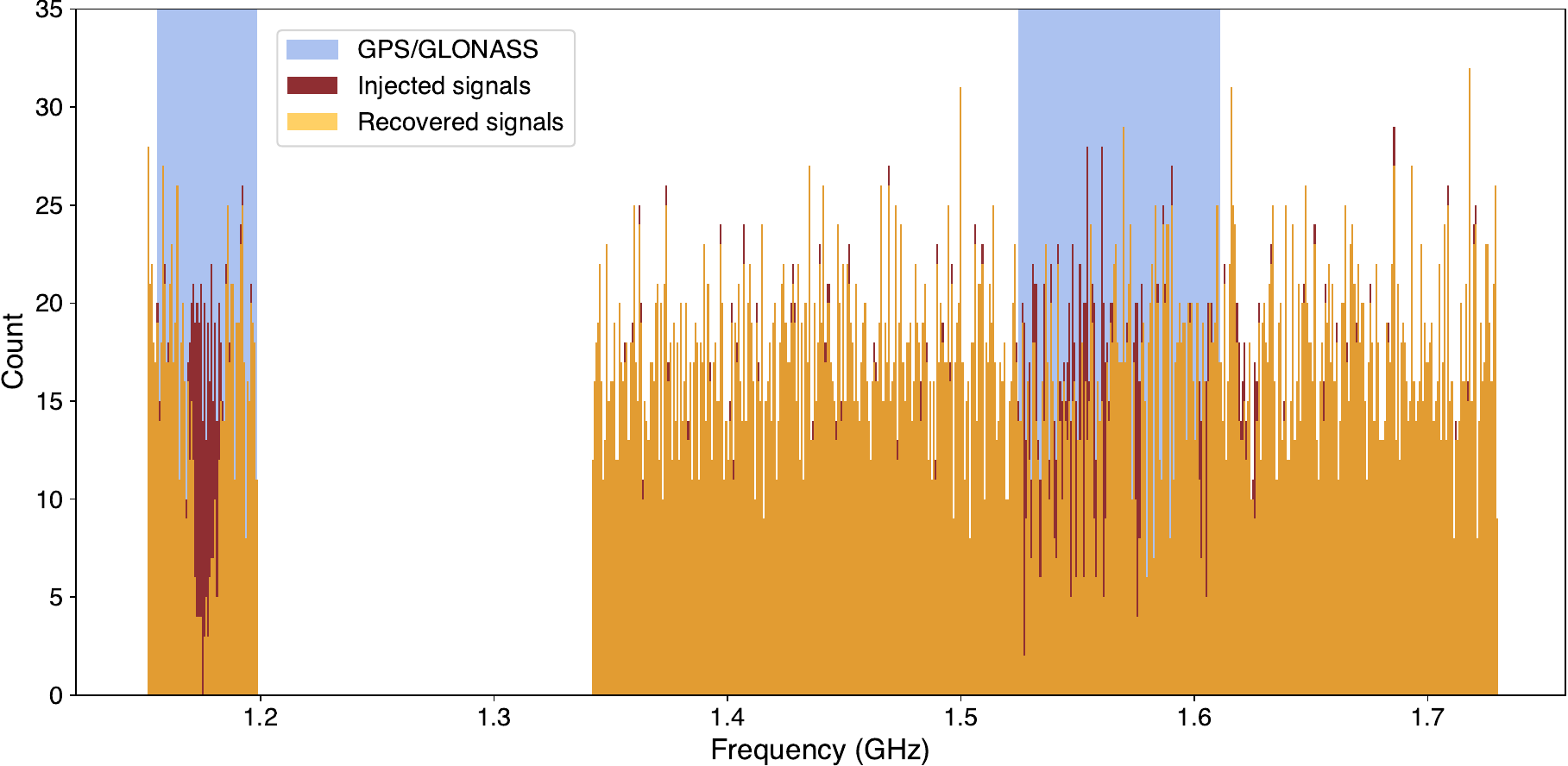}
\caption{Distribution of injected signals and signals recovered by the
  UCLA SETI pipeline as a function of frequency.  Blue bands indicated
  the operating bands of GPS and GLONASS satellites, where recovery
  rates are markedly lower.  Frequencies corresponding to a notch
  filter at the GBT (1.2 -- 1.3412 GHz) are excluded.  }
  \label{fig-ir}
\end{figure}

\begin{figure}[ht!]
  \begin{tabular}{cc}
    \includegraphics[width=0.45\textwidth]{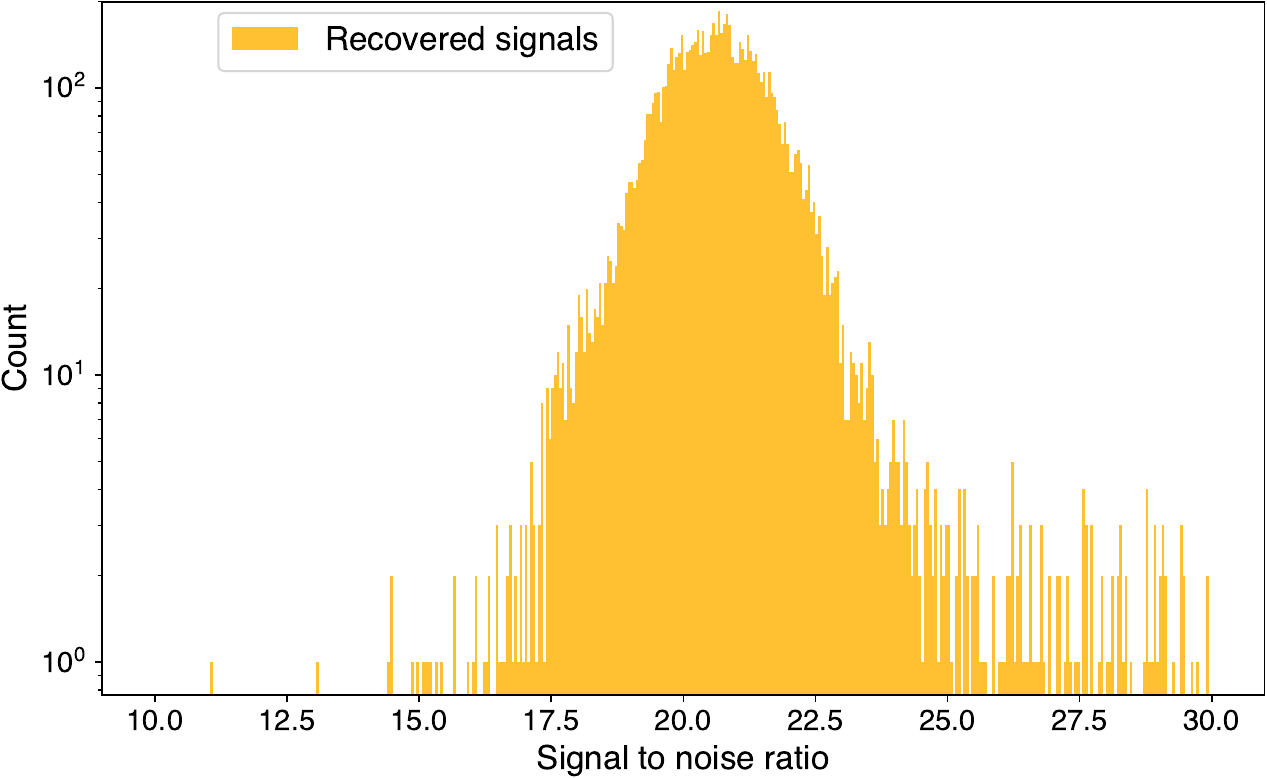} &
    \includegraphics[width=0.45\textwidth]{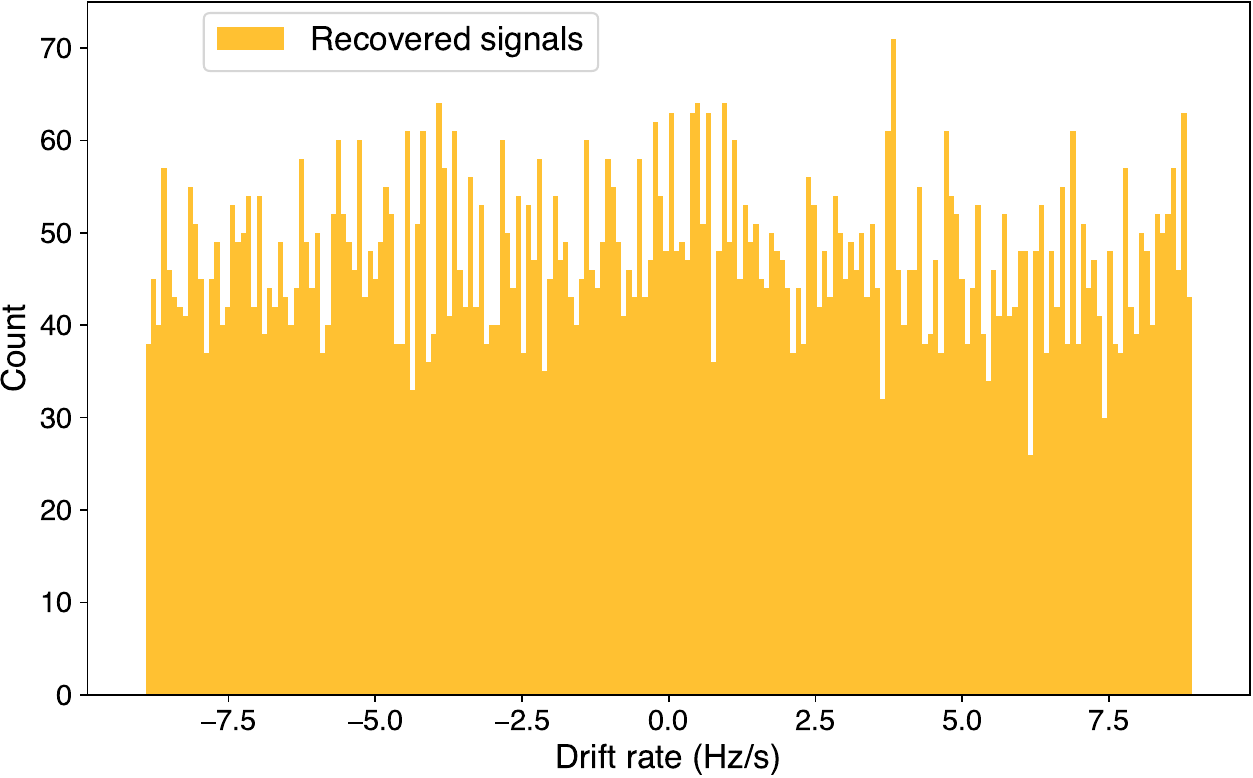}
  \end{tabular}
    \caption{Distributions of the recovered S/Ns (Left) and drift rates (Right)
  for the UCLA SETI pipeline.  }       
  \label{fig-ucla}
\end{figure}

The UCLA SETI pipeline recovered
9400 signals out of 10,000 injections, yielding an end-to-end pipeline
efficiency for narrowband chirp signals of
94\%.  When regions of dense RFI were excluded, the UCLA SETI pipeline
recovered
6716 signals out of 6807 injections,
for an improved
recovery rate of
98.7\% (Figure~\ref{fig-ir}).  The distributions of recovered S/Ns and
drift rates match those of the injected population
(Figure~\ref{fig-ucla}).  Signals that were not recovered are usually
found near the bandpass edges, where the bandpass response and
correction may be less than ideal, or intersect other signals in
time-frequency space.

\begin{figure}[ht!]
  \begin{tabular}{cc}
    \includegraphics[width=0.45\textwidth]{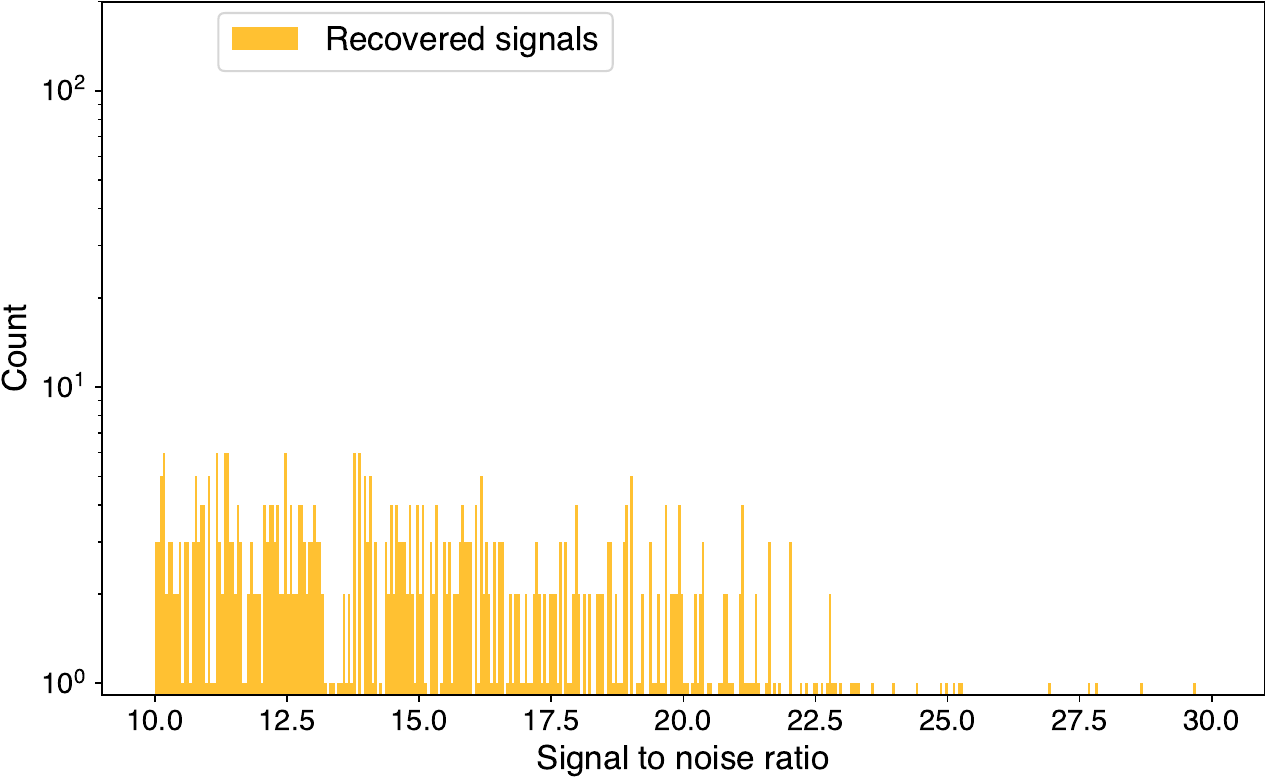} &
    \includegraphics[width=0.45\textwidth]{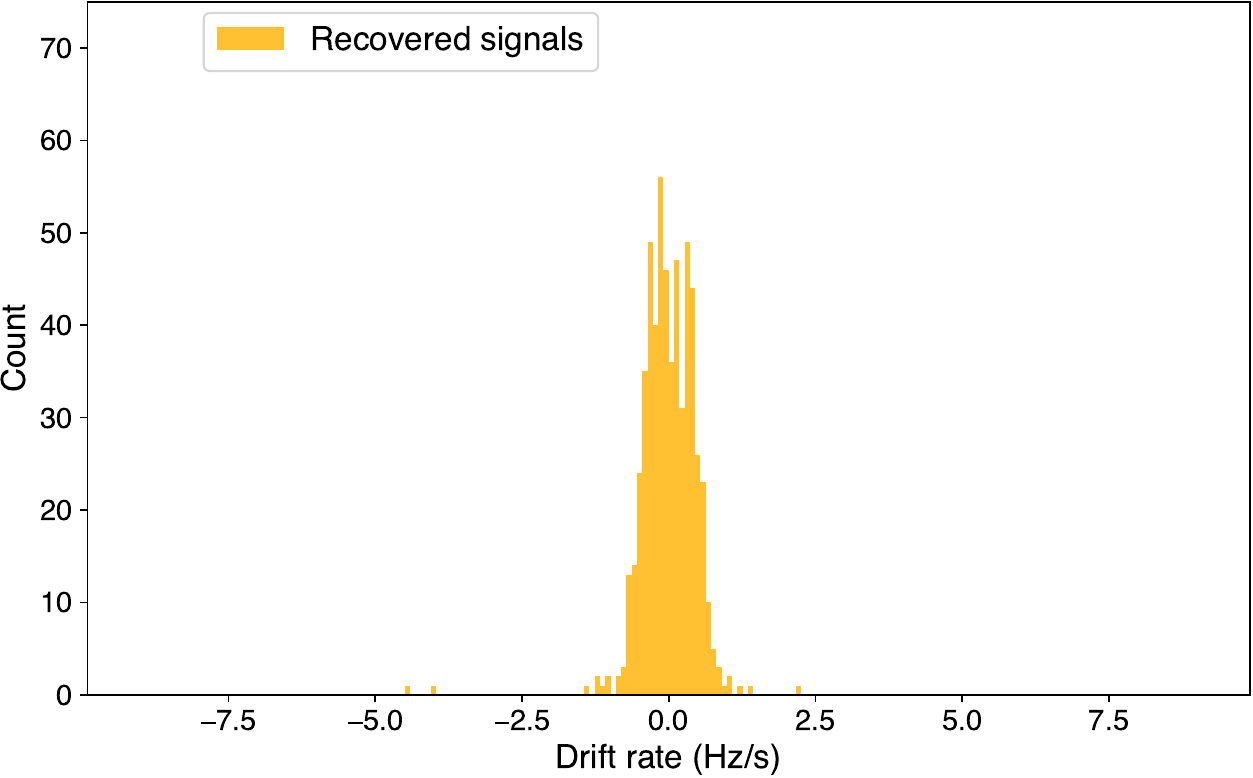}
  \end{tabular}
    \caption{Distributions of the recovered S/Ns (Left) and drift rates
      (Right) for a process that imitates the BL pipeline, i.e.,
      incoherent averaging of the spectra with $N_{\rm INC\_SUMS}=51$ followed by signal detection with 
      turboSETI.}
  \label{fig-bl}      
\end{figure}

\begin{figure}[ht!]
  \begin{center}
    \includegraphics[width=0.55\textwidth]{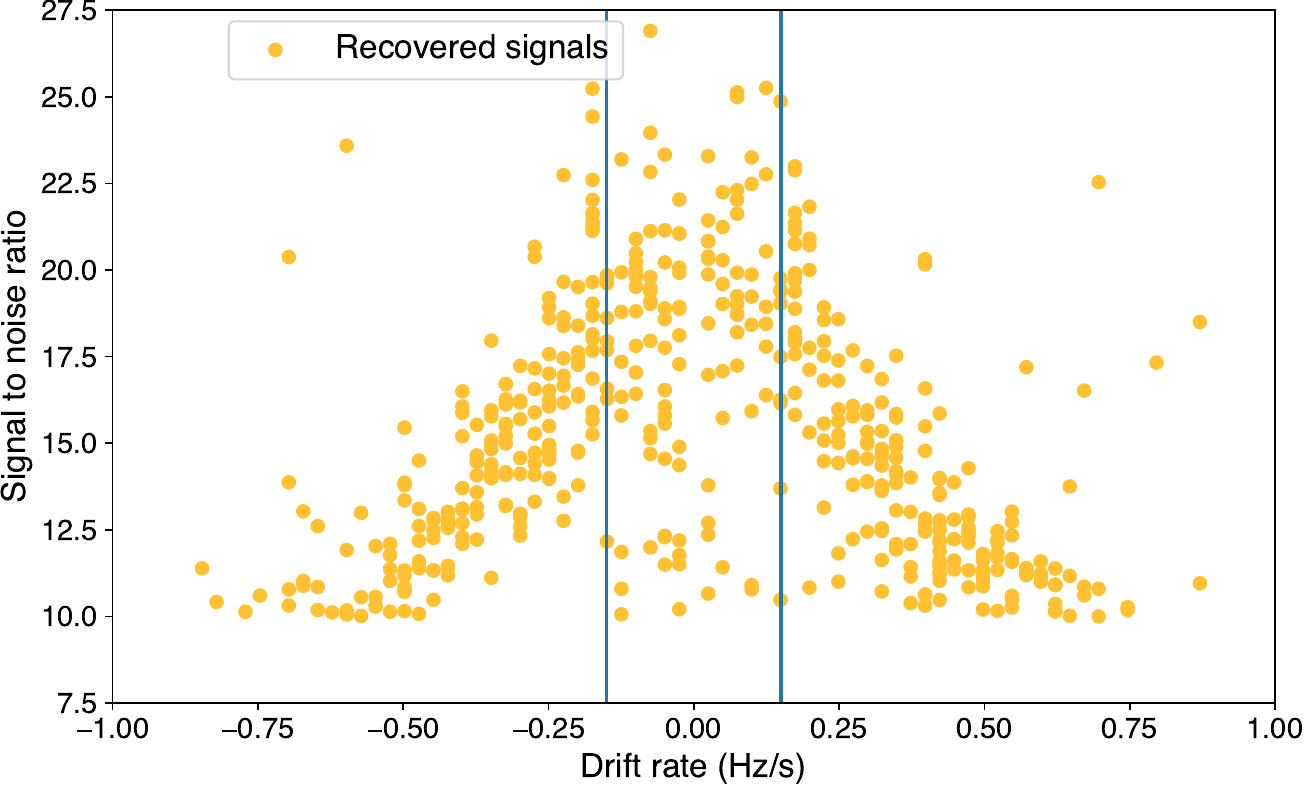}
  \end{center}
    \caption{Distribution of the recovered S/Ns as a function of drift rate
      (0--1 \Hzs only) for a process that imitates the BL pipeline.
      Only the central region within $\pm$0.15 \Hzs (blue lines) is
      free of Doppler smearing.  The recovered S/N values are lower
      than the injected values because of Doppler smearing that
      worsens at larger drift rates, as quantified by smaller
      dechirping efficiencies at larger drift rates.  Dechirping
      calculations for a BL-like process had predicted a drop in S/N
      to about $\sim$62\% of nominal for \dfdt = 0.25~Hz/s, $\sim$31\%
      for \dfdt = 0.5~Hz/s, and $\sim$16\% for \dfdt = 1~Hz/s
      \citep{marg21setiothers}, which is roughly consistent with what
      is observed.}
  \label{fig-zvsdfdt}      
\end{figure}

A process designed to imitate the BL pipeline recovered a much smaller
fraction of the injections.  Specifically, only
570 signals out of 10,000 injections were recovered, with S/N and
drift rate distributions that do not match the injected population and
illuminate the reasons for the poor performance (Figure~\ref{fig-bl}).
Almost all (99.1\%) signals recovered by the BL-like process have
drift rates within $\pm 1$ \Hzsns.  Injections with larger drift rates
are rarely recovered.  This result is entirely consistent with the
theoretical expectation of low dechirping efficiency for high drift
rate signals observed in incoherently summed power spectra.  In this
situation, the signal power gets smeared across multiple frequency
resolution cells because of Doppler drift during the longer
integration times.  For BL incoherent sums of 51 spectra at $\sim$3 Hz
resolution, which extend the integration times from $\sim$0.3 s to
$\sim$17~s, drift rates that exceed $\pm 0.15$ \Hzs experience Doppler
smearing.  The dechirping efficiency falls rapidly, reaching 16\% for
drift rates of 1 \Hzs \citep{marg21setiothers}, making recovery of
signals at larger drift rates challenging.  The diminishing
performance as a function of drift rate is evident when plotting the
S/Ns of signals recovered by the BL-like process as a function of drift
rate (Figure~\ref{fig-zvsdfdt}).

Detailed signal counts are listed in Table~\ref{tab-efficiency}.
These counts provide reasonable estimates of the end-to-end pipeline
efficiency of radio SETI pipelines.  The efficiency of a BL-like
process is $\sim$5.7\% for drift rates within $\pm$8.88 \Hzsns.
Because injected signals have uniformly distributed drift rates and
because all recovered signals have drift rates within the $\pm$4
\Hzs range used by \citet{pric20} and \citet{gajj21}, we can
estimate the end-to-end pipeline efficiency of their searches at
$5.7\% \times 8.88/4 = 12.7\% $.  Likewise, we find $5.7\% \times
8.88/2 = 25.3\% $ for the work of \citet{enri17}, who sampled drift
rates within $\pm$2 \Hzsns.

\begin{table}[h!]
  \begin{tabular}{lrrrrr}
    & Number of hits & Number of hits & Candidate  & Actual  & Pipeline \\
    & prior to injection & after injection & matches & matches & efficiency\\
    \hline
UCLA SETI pipeline &    329,591 & 338,238  & 9634 & 9400 & 94.0\% \\ %
BL-like process &  7512    &  8226  &  714  & 570  & 5.7\%   
  \end{tabular}
  \caption{Efficiency of radio SETI pipelines quantified by the
    recovery rates of 10,000 artificial signal injections.  In the
    UCLA SETI pipeline, a hit is defined as a narrowband signal detection
    with S/N $\geq$ 10.  In the BL pipeline, a hit has the additional
    requirement of a minimum distance ($\sim$ kHz) from previously
    recorded hits.  The factor of $\sim$50 difference in number of
    hits between UCLA and BL for the same data set and drift rate
    range has been previously documented and is understood primarily
    as the result of differences in dechirping efficiency and
    definition of a hit \citep{marg21setiothers}.  Because injected
    signals may replace one or more previously detected signals, the
    number of candidate matches after injection is not simply the
    difference in number of hits prior to and after injection.  Actual
    matches are defined as having both a recovered frequency
    within $\pm$6 Hz of the injected frequency and a drift rate within
    $\pm$0.05 \Hzs of the injected drift rate.  Although turboSETI
    in its debug mode can be coerced to reveal additional hits beyond
    its nominal hits, the number of actual matches in debug mode
    remains low at 670 recoveries out of 10,000 injections.  }
  \label{tab-efficiency}
\end{table}

\subsection{Native Detections of Machine Learning Candidates by the UCLA SETI Pipeline}
The UCLA SETI pipeline successfully detected MLc3, MLc4, MLc5, MLc7,
and MLc8 without invoking our own ML algorithms \citep{pinc22}.  We
did not attempt to detect MLc1, MLc2, and MLc6 because the drift rates
reported by \citet{ma23} for these signals (1.11 \Hzsns, 0.44 \Hzsns,
0.18 \Hzsns) exceed the nominal range of the O($N \log N$) tree
algorithm given the incoherent summing of 51 consecutive spectra in BL
HSR data products.  Based on the characteristics and appearance of the
signals, we predict that if the raw voltage data had been preserved,
we could have recovered MLc1, MLc2, and MLc6 by processing the data without incoherent summing.

The characteristics of the signals detected by the UCLA SETI pipeline
generally match those of the ML detections well (Table~\ref{tab-mlc}).
The magnitudes of the drift rates match, but the signs differ, which
we attribute to an error in \citet{ma23}'s report because our values
are consistent with the signal slopes in their supplemental figures.

\begin{table}[h]
  \begin{center}
\begin{tabular}{llrrrrrrrrr}
  ID &  Target&  Band Freq$_{\rm \,Ma}$ &  Freq$_{\rm \,UCLA}$ &  Offset & MJD$_{\rm Ma}$ &  MJD$_{\rm UCLA}$ &  DR$_{\rm Ma}$ & DR$_{\rm UCLA}$ &  S/N$_{\rm Ma}$ & S/N$_{\rm UCLA}$\\
     &  (HIP)      &     (Hz)     & (Hz)             & (Hz)    & (days)          &  (days)        & (\Hzs)        & (\Hzs)        &                & \\
\hline
MLc1  &  13402   & 1,188,539,231  &  N/A           &  N/A   &  57541.68902  & 57541.6890  &  +1.11  &   N/A   &    6.53  & N/A\\ 
MLc2  &  118212  & 1,347,862,244  &  N/A           &  N/A   &  57752.78580  & 57752.9095  &  -0.44  &   N/A   &   16.38  & N/A\\
MLc3  &  62207   & 1,351,625,410  &  1,351,623,638 &  -1772 &  57543.08647  & 57543.1000  &  -0.05  &  +0.049 &   57.52  &  80.31\\
MLc4  &  54677   & 1,372,987,594  &  1,372,984,455 &  -3139 &  57517.08789  & 57517.1017  &  -0.11  &  +0.11  &   30.20  &  41.71\\
MLc5  &  54677   & 1,376,988,694  &  1,376,984,409 &  -4285 &  57517.09628  & 57517.1017  &  -0.11  &  +0.108 &   44.58  &  63.50\\
MLc6  &  56802   & 1,435,940,307  &  N/A           &  N/A   &  57522.13197  & 57522.1527  &  -0.18  &   N/A   &   39.61  & N/A  \\
MLc7  &  13402   & 1,487,482,046  &  1,487,476,704 &  -5342 &  57544.51645  & 57544.5977  &  +0.10  &  -0.069 &  129.16  & 113.14\\
MLc8  &  62207   & 1,724,972,561  &  1,724,970,630 &  -1931 &  57543.10165  & 57543.1000  &  -0.126 &  +0.138 &   34.09  &  19.85\\
\end{tabular}
\caption{Characteristics of the top eight signals of interest identified by \citet{ma23}'s ML model and corresponding detections by the UCLA SETI pipeline.  MLc1, MLc2, and MLc6 have frequency drift rates beyond the nominal range of the O($N \log N$) tree algorithm and we did not attempt to detect them.  Columns 3, 6, 8, and 10 with subscripts ``Ma'' indicate the band frequency, start epoch, frequency drift rate, and S/N as reported by \citet{ma23}.  Columns 4, 5, 7, 9, and 11 with subscripts ``UCLA'' are corresponding UCLA SETI results.  \citet{ma23} did not report the frequencies of the signals but rather the center frequencies of the bands in which the signals were identified.  We report the actual frequencies of the signals at the beginning of each scan (column 4) and the frequency offsets (column 5) from the band centers.  The modified Julian dates (MJDs) reported by \citet{ma23} are erroneous except for the first one.  We provided the correct values (column 7).  We found frequency drift rates (DR) (column 9) that are opposite in sign to those reported by \citet{ma23} -- our values are consistent with the signal slopes in their supplemental figures.  S/N values differ because of algorithmic differences in outlier rejection when computing the standard deviation of the noise.}   
  \end{center}
  \label{tab-mlc}
  \end{table}

\begin{figure}[h!]
  \begin{tabular}{ccc}
  \includegraphics[width=0.3\textwidth]{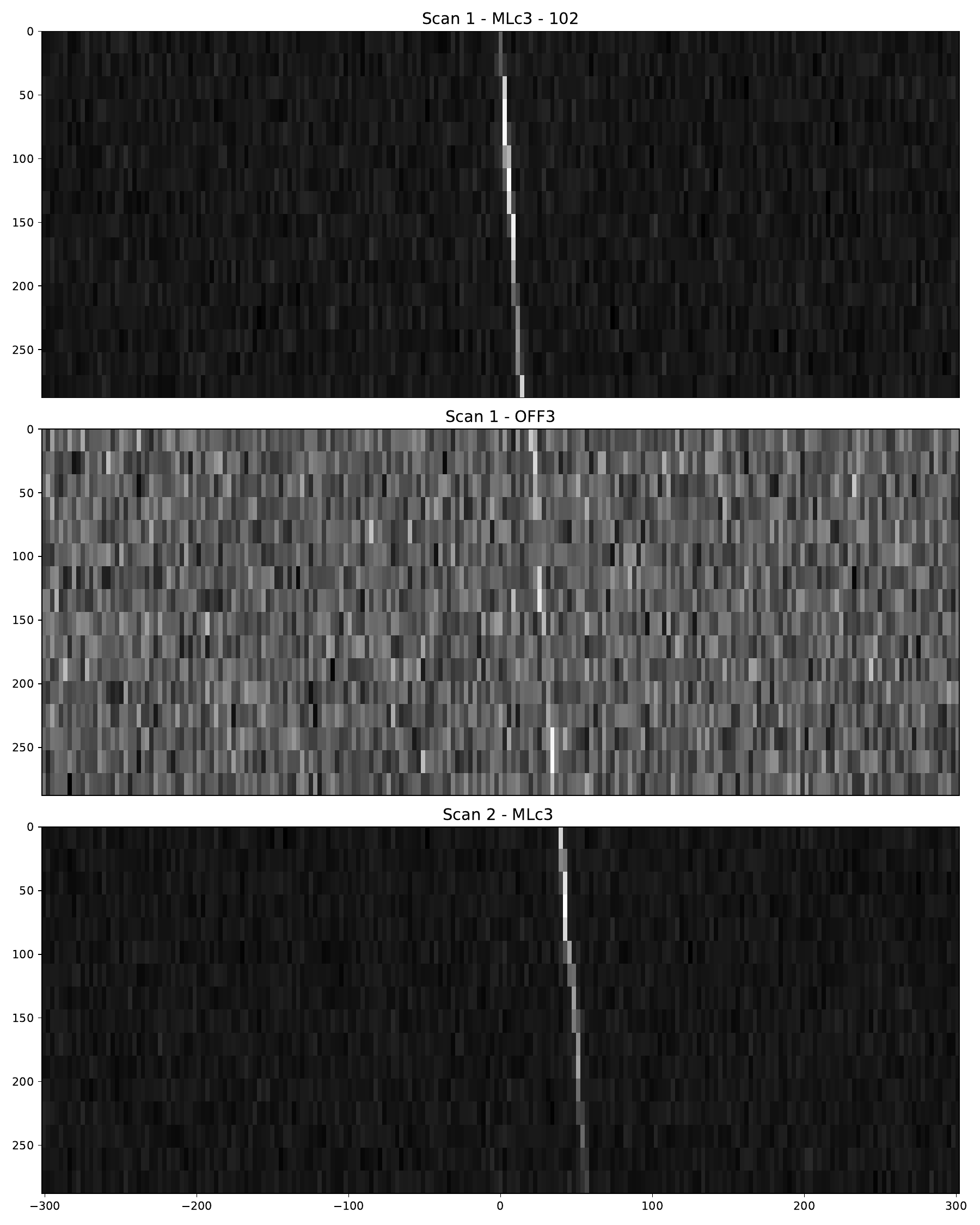} &
  \includegraphics[width=0.3\textwidth]{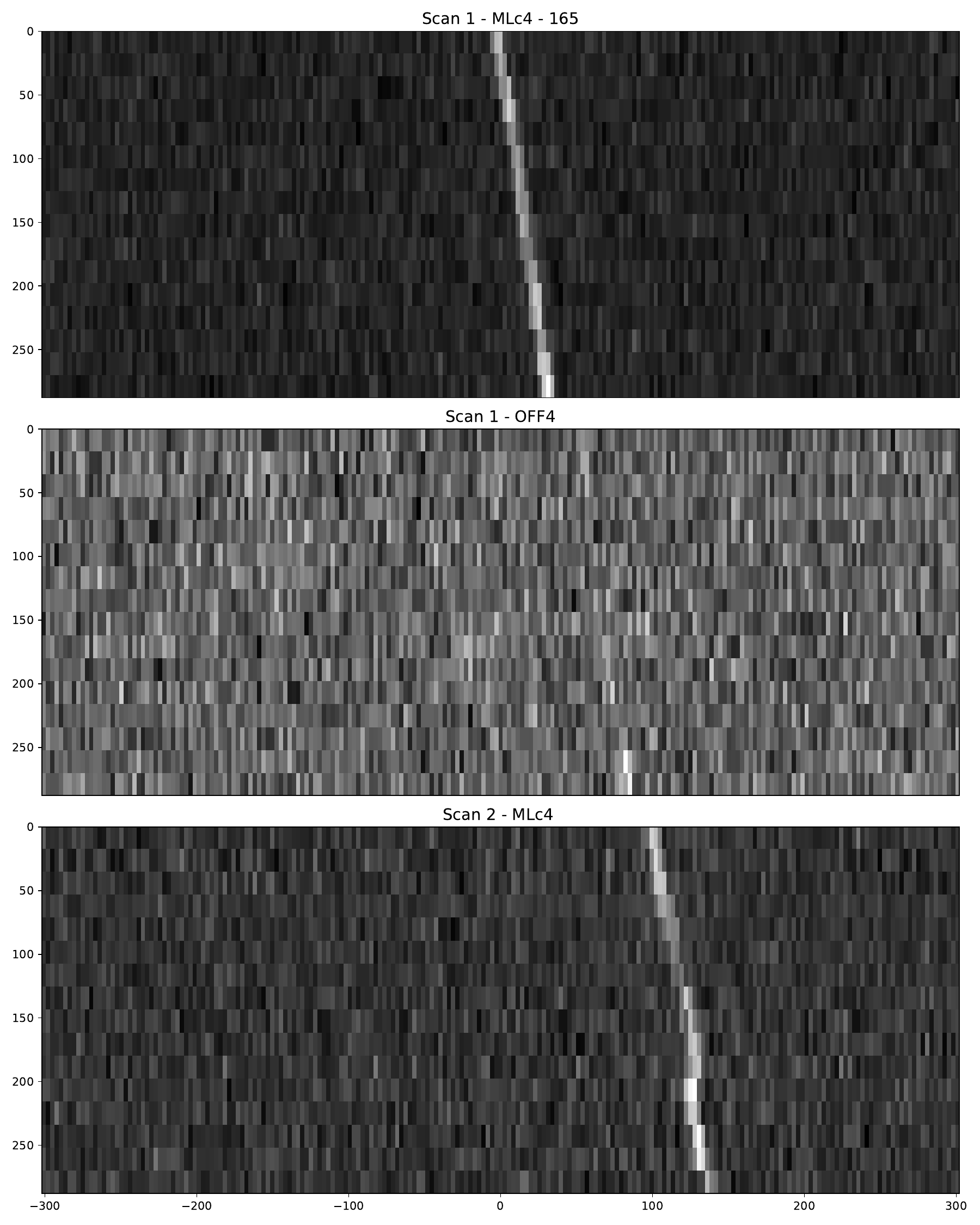} &
  \includegraphics[width=0.3\textwidth]{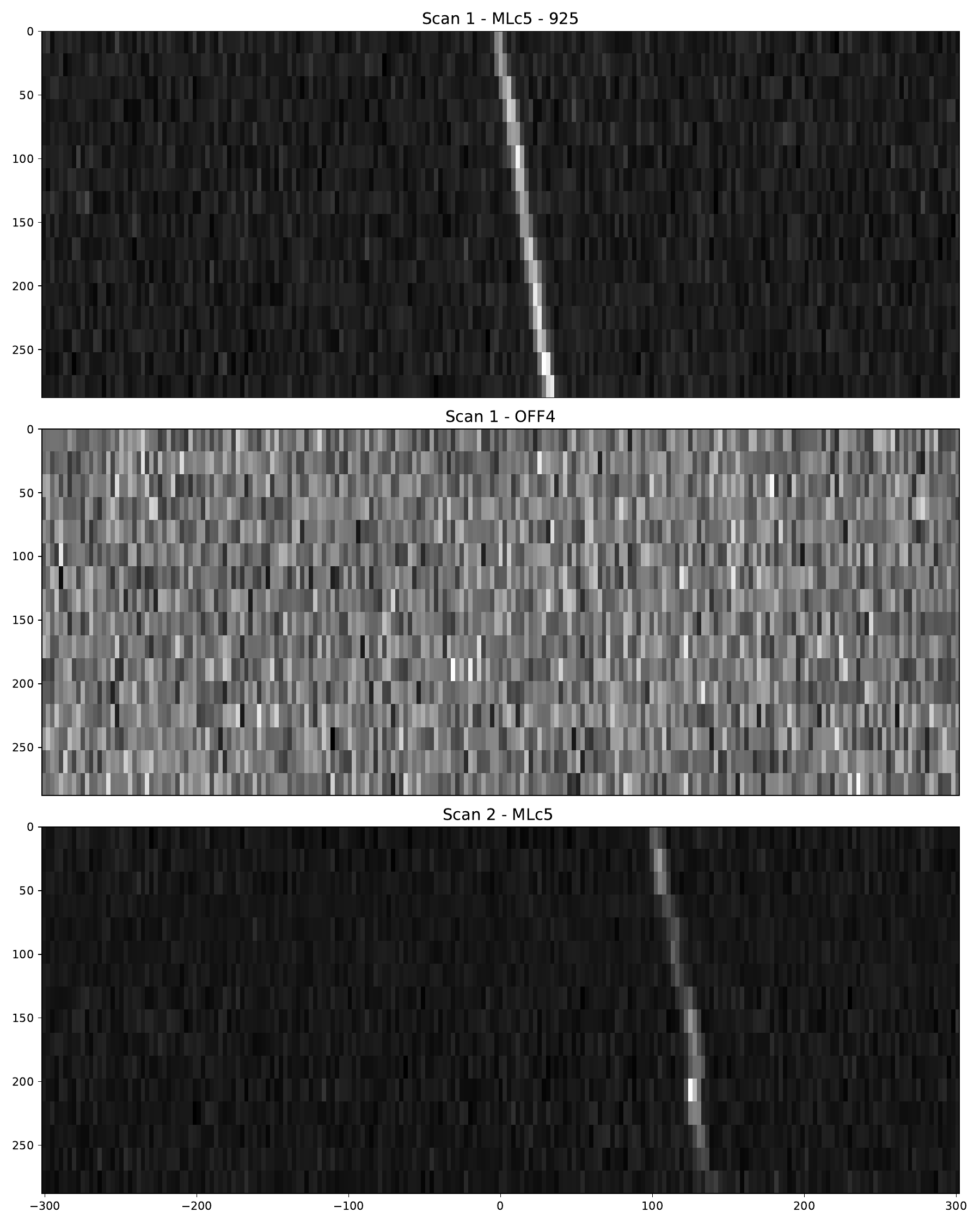}
  \end{tabular}
\caption{(Left) Dynamic spectra showing the ON-OFF-ON scans corresponding to MLc3 as detected by the UCLA SETI pipeline.  The detection of the signal in the OFF scan indicates that the signal can be immediately identified as RFI.  (Center) Same data for MLc4, which was identified by the UCLA SETI pipeline as a candidate worthy of visual inspection. (Right) Same data for MLc5, which was identified by the UCLA SETI pipeline as a candidate worthy of visual inspection.   }
  \label{fig-mlc}
\end{figure}

MLc3 was detected by the UCLA SETI pipeline but correctly identified 
automatically as RFI because the same signal is detected in the OFF
scan at S/N$\sim$12 (Figure ~\ref{fig-mlc}, Left).

MLc4 was detected by the UCLA SETI pipeline and identified as a
candidate warranting visual inspection.  Visual inspection clearly
reveals the presence of the signal in the OFF scan (Figure
~\ref{fig-mlc}, Center), indicating that this candidate is RFI.

MLc5 was detected by the UCLA SETI pipeline and identified as a
candidate warranting visual inspection.  Detection of the signal in
the OFF scan is less compelling (Figure ~\ref{fig-mlc}, Right), but
the similarity in signal morphology with MLc4 indicates that this
candidate is RFI.  The frequency spacing between MLc4 and MLc5 is
almost exactly 4 MHz, which suggests a common interferer.

MLc7 was detected by the UCLA SETI pipeline but correctly identified
automatically as RFI because the same signal is detected in the OFF
scan.

MLc8 was detected by the UCLA SETI pipeline and identified as a
candidate warranting visual inspection.  The signal is detected in the
OFF scan and therefore labeled as RFI.

In summary, none of the MLc signals detected in our work warrant
further examination.

\section{Discussion}
\label{sec-discussion}

\subsection{Figures of Merit}
\label{sec-fom}
The Drake Figure of Merit~\citep{drak84} provides an estimate of the
search volume of a SETI program that captures almost all essential
elements: frequency coverage, sky coverage, and sensitivity.  It is defined as
\begin{equation}
    {\rm DFM} = \frac{B\, \Omega}{S_{\rm det}^{3/2}},
\end{equation}
where $B$ is the total bandwidth examined, $\Omega$
is the fractional area of the sky covered,
and $S_{\rm det}$ is the minimum flux required for a detection.
Typical units are GHz m$^3$ W$^{-3/2}$ \citep[e.g.,][]{horo93}.

For transmitters of a given EIRP,
the $S_{\rm det}^{-3/2}$ factor is proportional to the total volume
that can be examined by a search with minimum detectable flux $S_{\rm
  det} (\propto {\rm EIRP} / 4\pi r_{\rm max}^2)$.  The fraction of this
volume that is actually sampled by the search is proportional to the
fraction $\Omega$ of a 4$\pi$ solid angle.  Multiple observations of the same patch of sky
with similar observing parameters can easily be accounted for by
rewriting $\Omega = \sum_i \Omega_i$, where the index $i$ represents
individual
{\rm observations}.
As such, $\Omega$ should be viewed as an effective solid angle and not a physical solid angle.
In this context, one observation is
  defined as a complete set of scans, e.g., two scans of source A for
  the UCLA SETI cadence.  The fraction of the entire radio spectrum
that is captured by the search is proportional to the bandwidth $B$.

In its original form, the DFM misses two essential elements.  First,
it assumes that pipelines are perfect with end-to-end pipeline
efficiencies of 100\%, whereas the efficiency of different programs
can vary by more than an order of magnitude (Table~\ref{tab-efficiency}).  Second, it ignores the
frequency drift rate coverage, i.e., range of line-of-sight accelerations
sampled in a search, whereas this range is an obvious indicator of the
thoroughness of the search.  We propose a modified DFM to address
these limitations:
\begin{equation}
    {\rm MDFM} = \eta_P \, \frac{\dot{v}_{\rm max}}{c} \frac{B\, \Omega}{S_{\rm det}^{3/2}},
\end{equation}
where $\eta_P$ is the end-to-end pipeline efficiency for the detection
of signals of interest and
$\dot{v}_{\rm max}/c$ is the maximum
fractional frequency drift rate (with respect to the center of the
band) considered in the search (Equation~\ref{eq-dfdt}).  We express
the latter in units of nHz and show MDFM values in units of Hz$^2$ m$^3$
W$^{-3/2}$.  We chose a metric that is linear in the range of
frequency drift rates examined because we cannot predict the
locations, sizes, or spins of preferred transmitter platforms.  In the
absence of reliable information, a uniform prior distribution for the
frequency drift rate seems reasonable.  One could design the
distribution to accommodate the majority of exoplanet settings, with
an upper limit of
26 nHz that accommodates
95\% of confirmed exoplanets with known semi-major axes and
orbital periods.

\begin{figure}[ht!]
  \begin{tabular}{cc}
    \includegraphics[width=0.45\textwidth]{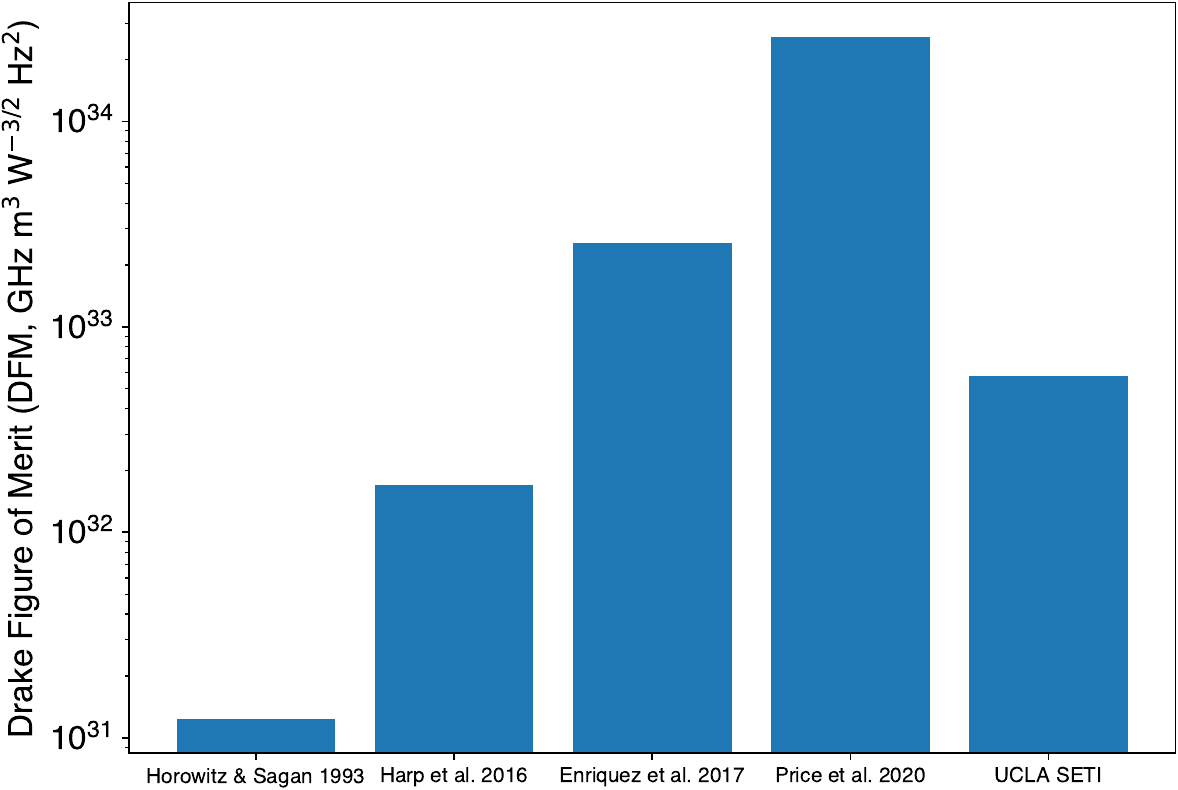} &
    \includegraphics[width=0.45\textwidth]{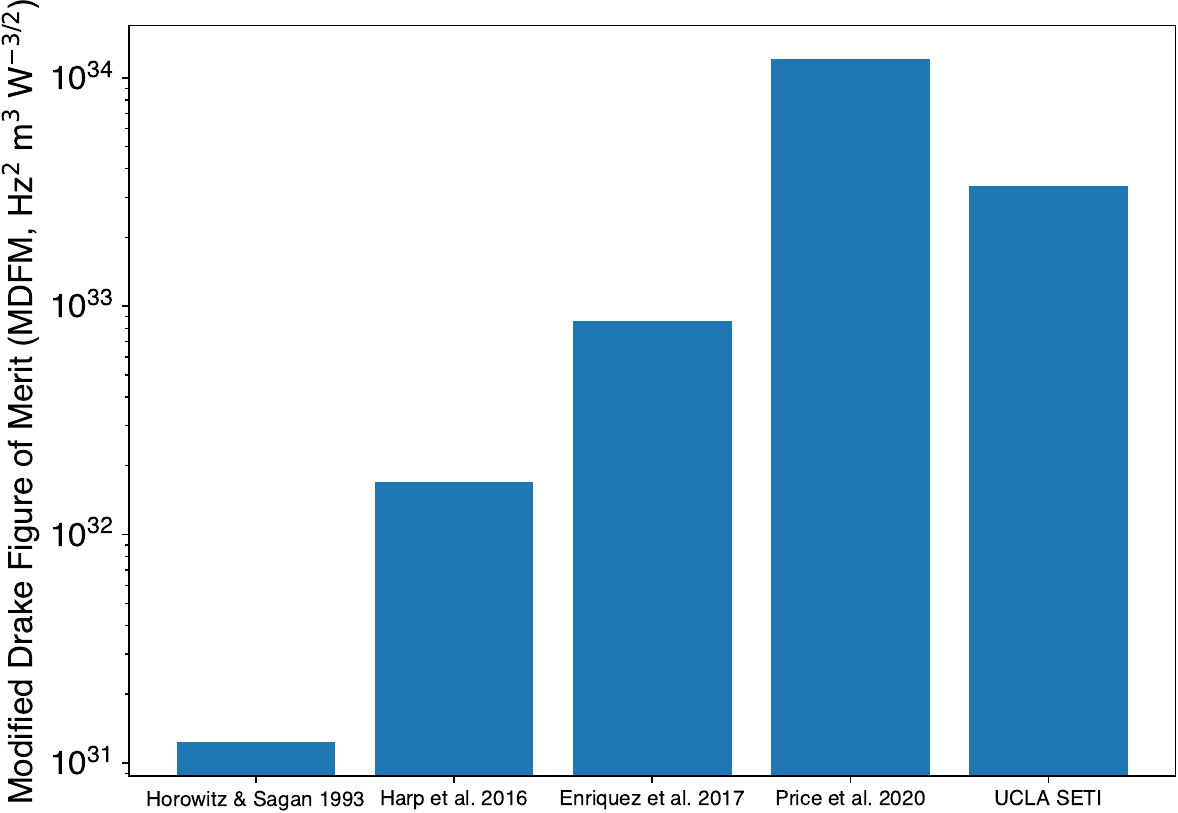}
  \end{tabular}
  \caption{
Search volume characteristics of select surveys.  The Drake
    Figure of Merit (DFM, Left) does not account for pipeline efficiency nor
    frequency drift rate coverage, but the Modified Drake
    Figure of Merit (MDFM, Right) does.
      }
  \label{fig-dfm}      
\end{figure}

Values of the DFM and MDFM metrics for the UCLA SETI search
are compared to those of select surveys in Table~\ref{tab-dfm} and
Figure~\ref{fig-dfm}.  We have assumed $\eta_P$ = 100\% for the
surveys of \citet{horo93} and \citet{harp16} and the estimates of
Section~\ref{sec-ir} for the surveys of \citet{enri17},
\citet{pric20}, and UCLA SETI.  The drift rate coverage of
\citet{horo93} is unlike those of modern surveys.  It is large but
samples only three distinct values (local standard of rest, galactic
barycenter, and cosmic microwave background rest frame).  We have
assumed a fractional drift rate of 1 nHz as a compromise, which is the
same value that \citet{harp16} used.
We found that the MDFM of the UCLA SETI search falls in between the
survey of 692 primary stars of \citet{enri17} and the survey of 1327
primary stars of \citet{pric20}.

\begin{deluxetable}{llllll}[h]
  \tablehead{
    &    \colhead{Horowitz \& Sagan 1993} & \colhead{Harp et al. 2016} & \colhead{Enriquez et al. 2017} & \colhead{Price et al. 2020} & \colhead{UCLA SETI}
    }
\startdata
  \hline
    Freq.\ coverage $B$ (GHz) & 4e-4 & variable$^a$    & 0.660   & variable$^b$  & 0.439 \\
    Sky fraction  $\Omega$        & 0.7  & 1.17e-3$^c$ & 2.88e-4 & 5.03e-4 & 4.91e-5 \\
    Sensitivity $S_{\rm det}$ (W/m$^2$) & 1700e-26$^d$ & 260e-26$^e$ &  17.7e-26$^f$ &  variable$^g$ &  11.3e-26$^h$\\
    Pipeline efficiency $\eta_P$  & 100\%& 100\%    & 25.3\% & 12.7\%  & 94.0\%\\
    Drift rate coverage (nHz)& 1    &  1       & 1.33 & 2.66 & 6.24         \\
    \hline
    DFM  (GHz m$^3$ W$^{-3/2}$)                     & 1.23e+31 & 1.70e+32 & 2.56e+33 & 2.59e+34 & 5.74e+32\\
    MDFM (Hz$^2$ m$^3$ W$^{-3/2}$)                    & 1.23e+31 & 1.70e+32 & 8.63e+32 & 1.21e+34 & 3.37e+33  
    \enddata
  \caption{Search volume characteristics of select surveys.  The Drake
    Figure of Merit (DFM) does not account for pipeline efficiency nor
    frequency drift rate coverage, but the Modified Drake
    Figure of Merit (MDFM) does.}
  \tablenotetext{a}{We used 8 GHz for 65 stars, 2.04 GHz for 1959 stars, 0.337 GHz for 2822 stars, and 0.268 GHz for 7459 stars \citep{harp16}.}
  \tablenotetext{b}{We used 0.66 GHz for GBT L band, 0.94 GHz for GBT S band, 0.85 GHz for Parkes 10 cm \citep{pric20}.}
  \tablenotetext{c}{Based on 3' $\times$ 6' synthesized beam. The solid angle appears to have been overestimated by a factor of 4 in \citet{enri17}.}
  \tablenotetext{d}{For S/N=30 in 20 s \citep{horo93}.}
  \tablenotetext{e}{For S/N=6.5 in 93 s \citep{harp16}.  We used a system temperature of 108 K, which is the average across all four bands.}
  \tablenotetext{f}{For S/N=25 in 300 s \citep{enri17}.}
  \tablenotetext{g}{For S/N=10 in 300 s \citep{pric20}.  We used 7.1e-26 for the GBT and 24.0e-26 for Parkes.}
  \tablenotetext{h}{For S/N=10 in 150 s.  This values takes the quantization efficiency $\eta_Q$ = 0.8825 into account.}
  \label{tab-dfm}
\end{deluxetable}

Another possible disadvantage of the DFM is that it assumes a uniform
distribution of transmitters on the sky, whereas transmitters may be
preferentially located near stars, which are not uniformly
distributed.  At $\sim$1\% of the galactic scale, the assumption of
spatial uniformity holds reasonably well.  For instance, the Gaia
catalog of nearby (100~pc) stars is expected to be volume-complete for
all stars of spectral type earlier than M8 and shows a roughly uniform
spatial distribution of the 331,312 objects~\citep{smar21}.  At larger
distances, the assumption breaks down, especially for directions
perpendicular to
the plane of the galactic disk.  \citet{drak84} had
anticipated this problem by considering distances $<$ 1 kpc.

As our galactic models and star catalogs improve,
we can refine the MDFM by replacing the physical volume covered by a
search with the actual number of stars sampled by each observation,
assuming again that transmitters may be preferentially located near
stars.  Let us consider the number of stars $dn_*$ in an elemental volume of
sky
\begin{equation}
dn_* (r, \theta, \phi) = \rho_* (r, \theta, \phi) r^2 \sin{\theta} d\theta d\phi dr,
\end{equation}
where $\rho_*$ is the stellar density (number of stars per unit volume),
and $(r, \theta, \phi)$ describe spherical coordinates in a frame
centered at the solar system barycenter.  The figure of merit for
transmitters of a fiducial EIRP can then be written
\begin{equation}
  {\rm MDFM|_{\rm EIRP}} = \eta_P \, \frac{\dot{v}_{\rm max}}{c} \, B \, \sum_i n_{*,i} 
\end{equation}
where the number of stars in each
observation $i$ is extracted from a
catalog query that includes distance and angular bounds or computed
from a galactic or extragalactic model
\begin{equation}
n_{*,i} = \int_o^{r_{\rm max}} dr \iint_{\Omega_i} \rho_* (r, \theta, \phi)  r^2 \sin{\theta} d\theta d\phi, 
\label{eq-nstars}
\end{equation}
with $r_{\rm max} = \sqrt{{\rm EIRP} / (4 \pi S_{\rm det})}$ and
$\Omega_i$ is the full width half max (FWHM) solid angle of the
telescope beam.  Note that the quantization and dechirping
efficiencies are properly taken into account via $S_{\rm det}$ and,
therefore, $r_{\rm max}$.
Multiple observations of the same stars are
allowed in these expressions to account for the fact that repeated
observations are valuable.

\subsection{Transmitter Prevalence Calculations}
\label{sec-prevalence}
We describe a formalism to calculate upper bounds on the
prevalence of civilizations operating transmitters detectable in SETI surveys.
Our calculation presupposes that
the observed stars form a representative sample of the population of
stars in the relevant search volume.

We write the total number of observed stars for transmitters of
a fiducial EIRP
\begin{equation}
  n_{\rm obs}|_{\rm EIRP} = \sum_i n_{*,i},
\end{equation}
where the number of stars is calculated as in Section~\ref{sec-fom}.
When including all stars within the solid angle defined by the antenna
beam FWHM, the EIRP ought to be augmented from its nominal value to
account for emissions detected in off-axis directions.  However, we
ignore this small correction, which is at most a factor of 2 at the
antenna beam's half maximum.

We consider the fraction $f_{\rm TX}$  of stars in the observed sample that
host a detectable transmitter, such that the number of detectable
transmitters in the observed sample is $n_{\rm TX} = f_{\rm TX} \times
n_{\rm obs}|_{\rm EIRP}$. If the observed sample is
representative of the entire search volume, $f_{\rm TX}$ can be used as an estimate that applies
to the entire search volume.
We wish to place an upper limit on $f_{\rm TX}$ on the basis of our
observations and the fact that we did not detect a technosignature.

We acknowledge the fact that SETI pipelines are not 100\% efficient.
For each observation of a detectable transmitter, the probability of
success for detection of the transmitter is $< 100\%$.  For data
analysis pipelines that have been characterized with an injection and
recovery analysis, we can set this probability to $\eta_P$, the
end-to-end pipeline efficiency.

We also acknowledge that a transmitter may not be detectable at all
times by considering the duty cycle $D$ of the transmitter, i.e., the
fraction of time that the transmitter is beaming in Earth's direction.

We write the probability of detecting a transmitter in each
observation of a star as 
\begin{equation}
p = f_{\rm TX} \eta_P D.
\end{equation}
We consider the result of our observations as the result of $n=n_{\rm
  obs}|_{\rm EIRP}$ independent trials, each with the same probability
of success $p$.  The number of successes in such an experiment is
given by the binomial distribution $B(n, p)$.

We determine the largest possible value of $f_{\rm TX}$ that is
consistent with obtaining zero successes in $n$ attempts at a confidence
level CL.  This value is obtained by solving
\begin{equation}
  (1-p)^{n} = 1 - {\rm CL},
  \label{eq-CL}
\end{equation}
i.e.,
\begin{equation}
  f_{\rm TX}^{u} = \frac{1 - (1 - {\rm CL})^{1/n}}{\eta_P D},
\label{eq-solve}
\end{equation}
where we have labeled the superscript u to denote the upper limit.
At the 95\% confidence level and for $\eta_P \simeq 1$, $D \simeq 1$, and $n > 20$,
this result is well approximated by the ``rule of three'' \citep{jova97}:
$f_{\rm TX}^{u} \simeq 3/n$.
This rule was derived but not named as such in the Cyclops Report~\citep[][p.\ 53]{cyclops}.

Our signal injection and recovery analysis indicates that the UCLA
SETI pipeline would have at most a 94.0--98.7\% probability of
detecting a narrowband technosignature in any given observation of a
star hosting a detectable transmitter.  If the transmission
frequencies are uniformly distributed in the range 1.15--1.73 GHz, the
probability is closer to 94.0\%.  If the transmission frequencies
happen to fall among radio astronomy protected bands or regions where
RFI is less severe, the probability is closer to 98.7\%.  We 
evaluate upper limits in the conservative case with $\eta_p=94\%$.

For this survey with 62 observations and the fiducial EIRP of 0.62 %
Arecibos
(1.35 $\times$ 10$^{13}$ W)
corresponding to detectability up
to 100~pc, we find $N_{\rm obs}|_{\rm EIRP} = 47$ in the Gaia
catalog of nearby stars~\citep{smar21}.
In this observed sample, the maximum transmitter fraction that is
compatible with our nondetection at a confidence level CL=95\% is
6.6\%, assuming a transmitter duty cycle of 100\%.  If this result is
generalizable to the entire search volume, there are at most 6.6\% of
the 331,312 stars within 100~pc that host a transmitter detectable in
our survey.
If we consider a fiducial EIRP=$5.08\times 10^{16}$ W that enables
detection of transmitters around any of the 10,230 observed stars
located within 20,000 ly, we find that the fraction of stars with
detectable transmitters is at most $3 \times 10^{-4}$.
This limit is more
stringent than those published by \citet{wlod20}, considering the necessary
revisions to their estimates (Section~\ref{sec-revisions}).

\begin{figure}[ht!]
  \begin{center}
    \includegraphics[width=0.55\textwidth]{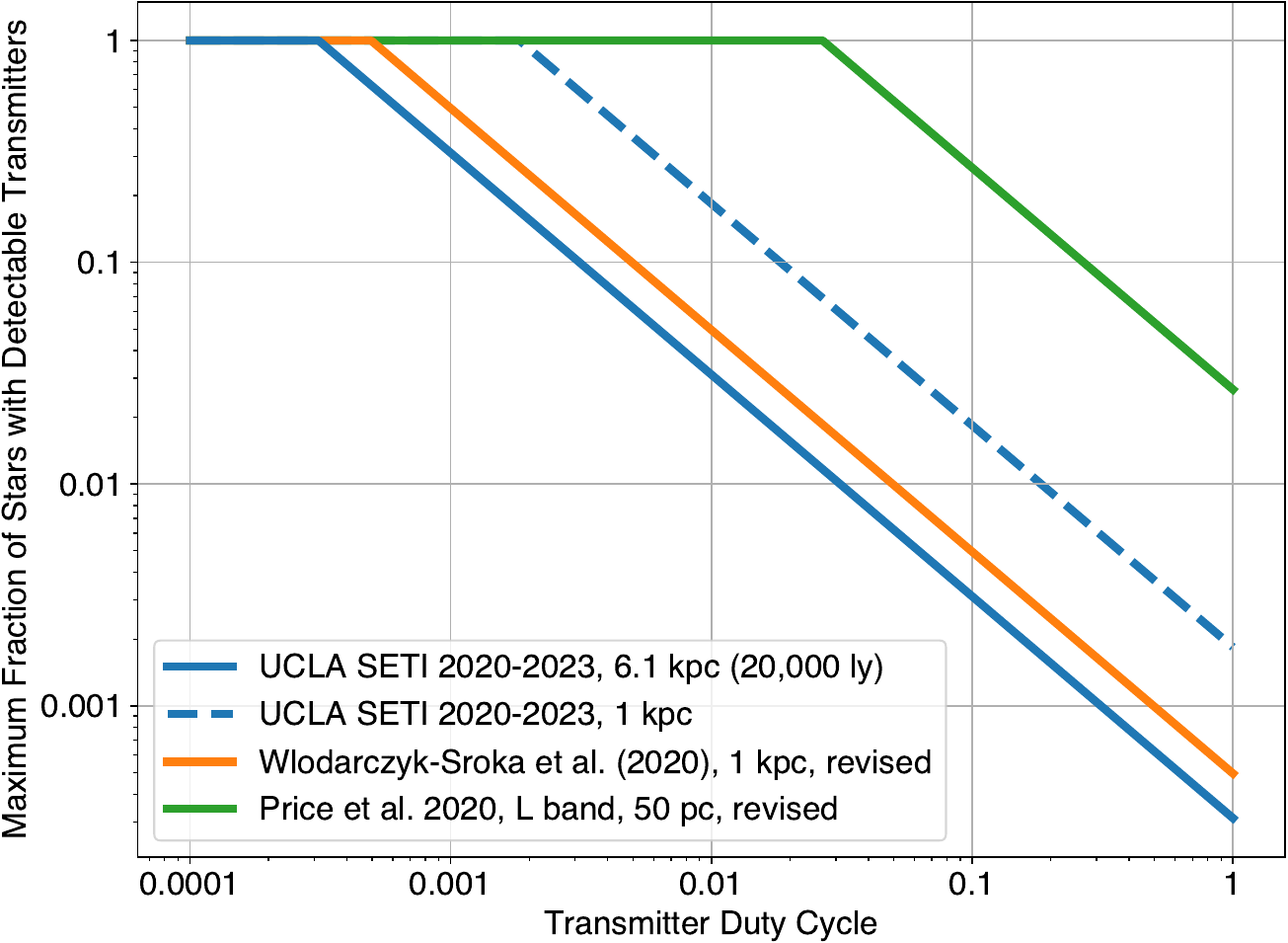}
  \end{center}
\caption{Upper limits on transmitter prevalence (95\% CL) as a function of transmitter duty cycle. The solid blue line shows the limit calculated for GBT observations of 10,230 stars located within 20,000 ly of the Sun and analysis with the UCLA SETI pipeline, which has a conservative 94\% probability of detecting detectable narrowband technosignatures (see text for detectability requirements). The dashed blue line shows the limit calculated for the subset of stars located within 1 kpc of the Sun. The solid and green orange lines show revisions to previously published limits (see Section~\ref{sec-revisions} for correction factors).}
  \label{fig-prevalence}      
\end{figure}

A detectable transmitter has the following sufficient characteristics:
(1) it emits in the frequency range 1.15--1.73 GHz (excluding the
range 1.20--1.34 GHz), (2) it has a line-of-sight acceleration with
respect to the GBT that results in a frequency drift rate within
$\pm8.86$ \Hzsns, and (3) it emits a fixed-frequency or chirp waveform
with bandwidth $<$ 3~Hz, 100\% duty cycle, and minimum EIRP as stated
above.  Characteristics (1) and (2) are necessary for detection, but
characteristics (3) are not.  For instance, we could detect more
complex or broader waveforms (e.g., pulsed waveforms, nonchirp
waveforms, or waveforms with $>$ 3~Hz bandwidth) provided that the
integrated power exceeded our detection threshold.  We could also
detect an intermittent transmitter provided that we observed at a
favorable time.
For transmitter duty cycles below 100\%, the upper limits on
transmitter prevalence are degraded (Figure~\ref{fig-prevalence}).

\subsection{Revisions to Published Estimates of Transmitter Prevalence}
\label{sec-revisions}
Many previous works
ignored the dechirping efficiency and provided estimates of SETI
search volumes or upper limits on the fraction of stars hosting
transmitters under the assumption that the end-to-end efficiency of
their pipeline was 100\% or 50\%.  Our results show that the
efficiency of a BL-like process is closer to 5.7\% for drift rates
within $\pm$8.88~\Hzsns, 12.7\% for drift rates within $\pm$4~\Hzsns,
and 25.3\% for drift rates within $\pm$2~\Hzsns, suggesting that the
published estimates of search volumes and transmitter limits in
previous works need revisions.  We propose revised estimates below.

We suggest that the statement of \citet{enri17} that ``fewer than
$\sim$0.1\% of the stellar systems within 50~pc possess these types of
transmitters'' ought to be rephrased as ``fewer than $\sim$1.7\% of
the stellar systems within 50~pc possess the types of transmitters
detectable in this search''.  Focusing on their 692 primary targets
only, which are stars within 50~pc, we can apply the formalism of
Section \ref{sec-prevalence} with $\eta_p=25.3\%$, which is
appropriate for frequency drift rates up to $\pm$2~\Hzs and BL data
products with 51-fold incoherent averaging.  We find that the upper
limit on $f_{\rm TX}$ is 1.7\% at the 95\% confidence level for their
692 primary targets, which is a more realistic upper limit on the
transmitter fraction.

\citet{pric20} described the results of a search of 882, 1005, and 189
primary targets at L band, S band, and 10~cm, respectively, with drift
rates up to $\pm$4~\Hzs in data products with 51-fold incoherent
averaging.  They used a detection probability $p=50\%$ in their
prevalence calculations, which yields
95\% CL prevalence estimates of 0.68\% at L band, 0.59\% at S band,
and 3.1\% at 10 cm. These values are approximately 50\% larger than
those published by \citet{pric20}, which are 0.45\%, 0.37\%, and
2.0\%, respectively, i.e., $\sim2/n$.  With the more realistic
pipeline efficiency of $\eta_p=12.7\%$, we find an upper limit of
$f_{\rm TX}=2.7\%$ at L band, suggesting that their upper limits must
be revised upwards by a factor of $\sim$6.

\citet{wlod20} improved the prevalence estimates by accounting for
stars captured in the telescope beam in addition to the primary
targets of \citet{enri17} and \citet{pric20}.  However, they combined
two data sets that are heterogeneous in their probabilities of
detection success, which makes revision of their estimates difficult.
Nevertheless, we can safely infer that their upper limit estimates must
be revised upwards by approximately an order of magnitude, i.e.,
between the correction factors of 6 and 17 that we identified for
these two surveys.  For instance, their estimate of the incidence of
star systems with detectable transmitters within 100~pc ought to be
revised from 0.061\% to $\sim$0.6\%.  Likewise, their estimate for
main-sequence stars within 1 kpc of 0.005\% ought to be revised to
0.05\%, slightly larger than our 0.03\% upper limit for stars within
20,000 ly.
We found 1732 stars within 1 kpc in our observed sample
(EIRP=$1.35\times10^{15}$ W), which leads to an upper limit of
transmitter prevalence of 0.18\% for these stars with our
observations.  For this 1 kpc search volume, the prevalence estimate
of \cite{wlod20}, with the correction presented here, is approximately
three times more stringent than ours but required approximately 100
times more telescope time (506.5 hr vs.\ 5.2 hr of GBT L-band on-source
integration time).
\citet{garr23}'s estimates for extragalactic transmitters require
similar revisions.

\citet{traa21} observed 28 TESS targets with 28 cadences at L, S, C,
and X band each, totaling $\sim$56 hr of on-source GBT time.  They used
a detection probability $p=50\%$ in their transmitter prevalence
calculations.  With their assumed pipeline efficiency, we find an
upper limit to the fraction of observed stars with transmitters of
20.3\% at the 95\% confidence level, which is not identical to their
estimate of 12.72\%.  Because they searched drift rates of up to
$\pm$4~\Hzs in data products with 51-fold incoherent averaging, a
pipeline efficiency of $\eta_p=12.7\%$ is more appropriate, and their
upper limits must be revised upwards by a factor of $\sim$6, leading
to a weak limit of $\sim$80\%.

\citet{fran22} searched 5, 17, 21, and 23 cadences of observations at
L, S, C, and X band, capturing a total of 61 TESS TOIs in transit with
$\sim$30 hr of GBT on-source time.  If the pipeline efficiency were
$\eta_p=50\%$, one might expect an upper limit on the fraction of
stars with transmitters of 90.0\%, 32.3\%, 26.6\%, and 24.4\%,
respectively, which differ from their values by more than $\sim$60\%.
However, the pipeline efficiency is closer to $\eta_p=12.7\%$, which
leads to an inability to place an upper limit at the 95\% confidence
level at L, S, and C band, and a weak limit of $\sim$96\% at X band.

\subsection{Optical SETI}
Shortly after \citet{cocc59} proposed
to search for interstellar communications in the radio part of the
spectrum, \citet{schw61} argued that similar searches could be
conducted in the optical part of the spectrum.  \citet{town83}
suggested that searches in both the microwave and infrared parts of
the spectrum were warranted.  In this section, we recognize the value
of searching at multiple wavelengths and briefly describe the results
of a few optical SETI initiatives, being mindful that a review of this
field is well beyond the scope of this work.
Radio and optical SETI searches are most likely to be successful if presumptive
civilizations operate beacons to emit distinctive signals, continuous or pulsed.  
Other information
carriers such as charged particles, massive particles, gravitational
waves, and neutrinos have been considered in the SETI context but are
deemed inferior to photons for communication purposes~\citep{hipp18}.

\citet{howa04} and \citet{ston05} described the results of multiyear
targeted searches of several thousand stars for nanosecond pulses
emitted by laser beacons.
\citet{tell17} adopted a different approach and searched for laser
emission lines in high-resolution spectra of 5600 nearby stars.
\citet{mair19} described the results of a search for near-infrared
pulses from 1280 celestial objects.

The prevalence calculations from these optical SETI surveys outperform
the radio SETI limits calculated for distances to 100 pc but not those
calculated for larger distances.  \citet{tell17} ruled out ``models of
the Milky Way in which over 0.1\% of warm, Earth-size planets harbor
technological civilizations that, intentionally or not, are beaming
optical lasers toward us.''  \citet{howa04} published transmitter
limits as a function of the transmitter repetition time.  The fraction
of stars with transmitting civilizations is also at most 0.1\% in
their work if one assumes repetition periods of $\sim3\times10^{4}$~h,
which may be reasonable considering that it would take us
approximately that long to beam a laser at all 331,312 stars within
100~pc with a 5 minute dwell time (equivalent duty cycle $D_{\rm opt}
\simeq 3 \times 10^{-6}$).  The upper limit on transmitter prevalence
that we obtained is $3 \times 10^{-4}$, but only for transmitters with
$D_{\rm radio} = 100\%$.

From the point of view of the transmitting civilization, the
requirements that an optical beacon be pointed at each one of tens of
billions of stars is considerably more onerous than the equivalent
requirement at radio wavelengths, where the entire sky can be covered
much faster.  For instance, the ratio of broadcast solid angles for
beamwidths of 8.4 arcminute (GBT-class telescope at L band) and 1
arcsec \citep[optical telescope, ][]{town83} is $2.5 \times 10^5$.
This ratio also dictates the ratio of duty cycles $D_{\rm radio} /
D_{\rm opt}$, which may justify the comparison with different values
of the duty cycles in the paragraph above.
To a civilization intent on broadcasting its existence, this intrinsic
advantage of radio beacons may eclipse other considerations.  However,
we are unable to anticipate the choices of presumptive civilizations,
and it behooves the SETI community to pursue a variety of search
modalities.

\FloatBarrier
\section{Conclusions}
\label{sec-conclusions}

Our observations of $\sim$11,680 stars and planetary systems with the
GBT resulted in $\sim$37 million narrowband detections, none of which
warranted reobservation.

A signal injection and recovery analysis of 10,000 chirp signals with
randomly selected frequencies and drift rates revealed that the UCLA
SETI pipeline recovers 94.0\% of the injections and 98.7\% of the
injections outside of regions of dense RFI.  Because the artificial
signals were injected in raw voltage data, these percentages represent
good estimates of the end-to-end pipeline efficiency for chirp
signals.

A process that simulates the BL pipeline recovers a much smaller
fraction of injections (5.7\%), which we attribute largely to Doppler
smearing of the signal that results from incoherent summing of 51
consecutive spectra.  The characteristics of the recovered signals
match the dechirping efficiency predictions of
\citet{marg21setiothers} and confirm that the dechirping efficiency is
an important factor that affects sensitivity, figure-of-merit, and
transmitter prevalence calculations.

We developed a formalism for the calculation of upper limits on
transmitter prevalence that take the end-to-end pipeline efficiency
and transmitter duty cycle into account.  We presented values
calculated at the 95\% confidence level for duty cycles of 100\% and
assumed that our observed sample is representative of the search volume.
On the basis of
our results and a Gaia survey of nearby (100~pc) stars,
we can state
that fewer than 6.6\% of the 331,312 stars within 100~pc host a
transmitter that is detectable in our survey (EIRP $> 1.35 \times
10^{13}$ W).
If we extend the search volume to 1~kpc, the limit becomes 0.18\%
(EIRP $> 1.35 \times 10^{15}$ W).  For stars located within 20,000 ly,
we found that the fraction of stars with detectable transmitters (EIRP
$>5.08\times 10^{16}$ W) is at most $3 \times 10^{-4}$.  Provided that
the frequency and frequency drift rate fall within our search bounds,
a sufficient condition for detection is the emission of a
fixed-frequency or chirp waveform with bandwidth $<$ 3~Hz, 100\% duty
cycle, and minimum EIRP as stated above.  We found that several
previously published prevalence estimates need revisions with
correction factors between 6 and 17.

We showed that the UCLA SETI pipeline can detect signals that had
escaped the BL pipeline and were identified with AI techniques by
\citet{ma23}.  In addition, we found that the AI detections are due to
RFI, either because our pipeline correctly and automatically
identified them as RFI, or because our usual visual inspection process
showed them to be RFI.

We developed an improved Drake Figure of Merit for SETI search volume
calculations that includes the pipeline efficiency and frequency drift
rate coverage of a search.  With this search volume metric, the UCLA
SETI search to date falls in between the survey of 692 primary stars of
\citet{enri17} and the survey of 1327 primary stars of \citet{pric20}.

UCLA SETI observations were designed, obtained, and analyzed by
$\sim$130 undergraduate and $\sim$20 graduate students who have taken
the annual SETI course since its first offering in 2016.
74 such students are coauthors of this work.
The SETI
course helps develop skills in astronomy, computer science, signal
processing, statistical analysis, and telecommunications.  Additional
information about the course is available at
\href{https://seti.ucla.edu}{https://seti.ucla.edu}.  UCLA SETI data are used in a citizen
science collaboration called ``Are we alone in the universe?'', which can be found at 
\href{http://arewealone.earth}{http://arewealone.earth}.

\clearpage
\begin{acknowledgments}
Funding for UCLA SETI was provided by The Queens Road Foundation,
Robert Meadow and Carrie Menkel-Meadow, Larry Lesyna, Michael
W. Thacher and Rhonda L. Rundle, Janet Marott, and other donors
(\url{https://seti.ucla.edu/donor-recognition}).  Part of this work was funded by NASA
Exoplanets Research Program grant 80NSSC21K0575.  We are grateful to
an anonymous referee for their review and for suggesting a section on
optical SETI.  We are grateful to the BL team for stimulating
discussions about search modes and data processing.  We thank the Tsay
Family, Smadar Gilboa, Marek Grzeskowiak, and Max Kopelevich for
providing an excellent computing environment in the Jim and Barbara
Tsay Computer Study Lab at UCLA.  We are grateful to Paul Demorest,
John Ford, Ron Maddalena, Toney Minter, Karen O’Neil, James Jackson,
and all the GBT support staff for enabling the GBT observations.  We
are grateful to Norma A. Contreras, Samuel R. Mason, Lou Baya Ould
Rouis, Taylor L.  Scott, and Nathanael Smith for assistance with the
data analysis and to Siavash Jalal for discussions about probability distributions.
The Green Bank Observatory is a
facility of the National Science Foundation operated under cooperative
agreement by Associated Universities, Inc.  This paper includes data
collected by the TESS mission. Funding for the TESS mission is
provided by the NASA's Science Mission Directorate.  This work has
made use of data from the European Space Agency (ESA) mission {\it
  Gaia},
processed by the {\it Gaia}
Data Processing and Analysis Consortium (DPAC).
This research has made use of the
\citet{exoarchive},
which is
operated by the California Institute of Technology, under contract
with NASA under the Exoplanet Exploration Program.
\end{acknowledgments}

\facilities{GBT, Exoplanet Archive}

\pagebreak
\appendix
\section{Sources}
\label{app-sources}

\begin{table}[h]
\begin{tabular}{rlrrrrrll}
\toprule \toprule TOI & Disp. & $R_p$ & Period & Insolation & $R_s$ &
Distance & RA & Dec \\ & & ($R_\earth$) & (days) & (Earth flux) &
($R_\sun$) & (pc) & (hh:mm:ss) & (dd:mm:ss) \\
\toprule
 469.01 &                 CP &                     3.55 &          13.63 &                           60.16 &                    1.01 &                  68.19 & 06:12:13.88 & -14:38:57.54 \\
 479.01 &                 KP &                    12.68 &           2.78 &                          615.47 &                    1.02 &                 194.55 & 06:04:21.53 &  -16:57:55.4 \\
 488.01 &                 CP &                     1.12 &           1.20 &                           58.26 &                    0.35 &                  27.36 & 08:02:22.47 &  03:20:13.79 \\
 536.01 &                 KP &                    15.40 &           9.24 &                          180.87 &                    1.30 &                 844.06 &  06:30:52.9 &  00:13:36.82 \\
 546.01 &                 KP &                    13.44 &           9.20 &                          203.95 &                    1.12 &                 726.41 & 06:48:46.71 & -00:40:22.03 \\
 561.01 &                 CP &                     2.74 &          10.78 &                           73.35 &                    0.84 &                  85.80 & 09:52:44.44 &  06:12:57.97 \\
 562.01 &                 CP &                     1.22 &           3.93 &                           14.71 &                    0.36 &                   9.44 & 09:36:01.79 & -21:39:54.23 \\
 571.01 &                 KP &                    12.92 &           4.64 &                          469.05 &                    1.41 &                 405.24 & 09:01:22.65 &   06:05:49.5 \\
 652.01 &                 CP &                     2.11 &           3.98 &                          464.96 &                    1.03 &                  45.68 & 09:56:29.64 & -24:05:57.07 \\
 969.01 &                 CP &                     3.65 &           1.82 &                          167.63 &                    0.82 &                  77.26 &  07:40:32.8 &  02:05:54.92 \\
1235.01 &                 CP &                     1.89 &           3.44 &                          134.67 &                    0.63 &                  39.63 & 10:08:52.38 &  69:16:35.83 \\
1243.01 &                 PC &                     4.49 &           4.66 &                            8.59 &                    0.49 &                  43.19 & 09:02:55.83 &   71:38:11.1 \\
1718.01 &                 PC &                     4.40 &           5.59 &                          219.06 &                    0.94 &                  52.30 & 07:28:04.33 &  30:19:18.24 \\
1726.01 &                 CP &                     2.24 &           7.11 &                          145.57 &                    0.90 &                  22.40 & 07:49:55.05 &  27:21:47.28 \\
1730.01 &                 PC &                     2.76 &           6.23 &                          114.27 &                    0.53 &                  35.69 &  07:11:27.8 &  48:19:40.56 \\
1732.01 &                 PC &                     2.55 &           4.12 &                           38.33 &                    0.63 &                  74.76 & 07:27:12.35 &  53:02:42.97 \\    
\toprule
1766.01 &                 KP &                    16.72 &           2.70 &                         1704.57 &                    1.61 &                 210.25 & 09:54:34.35 &  40:23:16.6 \\
1774.01 &                 CP &                     2.74 &          16.71 &                           73.76 &                    1.09 &                  53.97 & 09:52:38.86 & 35:06:39.63 \\
1775.01 &                 PC &                     8.70 &          10.24 &                           55.44 &                    0.84 &                 149.23 & 10:00:27.62 &  39:27:27.9 \\
1776.01 &                 PC &                     1.40 &           2.80 &                          560.33 &                    0.95 &                  44.65 & 10:59:06.55 & 40:59:01.39 \\
1779.01 &                 KP &                     9.93 &           1.88 &                           21.19 &                    0.31 &                  33.93 & 09:51:04.45 &  35:58:06.8 \\
1789.01 &                 CP &                    16.86 &           3.21 &                         3000.05 &                    2.26 &                 229.07 & 09:30:58.42 & 26:32:23.98 \\
1797.01 &                 CP &                     2.99 &           3.65 &                          283.72 &                    1.05 &                  82.34 & 10:51:06.41 & 25:38:27.83 \\
1799.01 &                 PC &                     1.63 &           7.09 &                          163.47 &                    0.96 &                  62.13 &  11:08:55.9 & 34:18:10.85 \\
1800.01 &                 KP &                    12.42 &           4.12 &                          459.54 &                    1.26 &                 277.28 & 11:25:05.98 & 41:01:40.87 \\
1801.01 &                 PC &                     1.99 &          10.64 &                           10.73 &                    0.55 &                  30.68 & 11:42:18.14 & 23:01:37.32 \\
1802.01 &                 PC &                     2.51 &          16.80 &                            5.92 &                    0.58 &                  60.69 & 10:57:01.28 & 24:52:56.42 \\
1803.01 &                 PC &                     4.22 &          12.89 &                           18.34 &                    0.69 &                 119.24 & 11:52:11.07 & 35:10:18.48 \\
1806.01 &                 PC &                     2.84 &          15.15 &                            2.15 &                    0.40 &                  55.52 & 11:04:28.36 & 30:27:30.87 \\
1821.01 &                 KP &                     2.43 &           9.49 &                           41.92 &                    0.77 &                  21.56 & 11:14:33.04 & 25:42:38.15 \\
1822.01 &                APC &                    14.57 &           9.61 &                          192.62 &                    1.71 &                 312.52 & 11:11:06.68 & 39:31:36.02 \\
1898.01 &                 PC &                     9.17 &          45.52 &                            8.34 &                    1.61 &                  79.67 & 09:38:13.27 & 23:32:48.29 \\    
\toprule
\end{tabular}
\caption{Characteristics of primary sources observed in 2020--2021.
  Columns show the TESS Object of Interest (TOI); the TESS Follow-up Observing Program Working Group disposition as of 2023 March 29 (Disp.), where PC is a planet candidate, CP is a confirmed planet, KP is a Kepler planet, and APC is an ambiguous planet candidate; the planet radius  $R_p$ in Earth radii; the orbital period in days; the insolation in Earth flux units; the radius of the host star $R_s$  in solar radii; the distance in parsecs; and the right ascension and declination of the source.  }
\label{tab-targets1}
\end{table}

\begin{table}[h]
\begin{tabular}{rlrrrrrll}
\toprule \toprule TOI & Disp. & $R_p$ & Period & Insolation & $R_s$ &
Distance & RA & Dec \\ & & ($R_\earth$) & (days) & (Earth flux) &
($R_\sun$) & (pc) & (hh:mm:ss) & (dd:mm:ss) \\
\toprule
1683.01 &                 PC &                     2.64 &           3.06 &                          163.06 &                    0.70 &                  51.19 & 04:23:55.12 & 27:49:20.53 \\
1685.01 &                 CP &                     1.32 &           0.67 &                          204.71 &                    0.46 &                  37.62 & 04:34:22.55 & 43:02:13.34 \\
1693.01 &                 CP &                     1.42 &           1.77 &                           57.02 &                    0.46 &                  30.79 &    06:01:14 & 34:46:23.13 \\
1696.01 &                 CP &                     3.17 &           2.50 &                           13.81 &                    0.28 &                  64.92 & 04:21:07.36 & 48:49:11.39 \\
1713.01 &                 PC &                     4.65 &           0.56 &                         3415.83 &                    0.95 &                 138.37 & 06:42:04.94 & 39:50:34.45 \\
1730.01 &                 PC &                     2.76 &           6.23 &                          114.27 &                    0.53 &                  35.69 &  07:11:27.8 & 48:19:40.56 \\
3772.01 &                 PC &                     7.32 &           4.17 &                          201.89 &                    0.87 &                 309.14 & 05:44:10.44 & 36:04:50.35 \\
3795.01 &                 PC &                     6.47 &           2.83 &                          462.81 &                    1.01 &                 439.84 & 06:34:55.79 & 49:40:35.67 \\
3800.01 &                 PC &                     5.74 &           1.67 &                         3970.53 &                    1.40 &                 598.89 & 06:53:06.26 &  39:07:56.2 \\
4596.01 &                 PC &                     2.72 &           4.12 &                          186.21 &                    0.98 &                  93.49 & 06:34:49.88 & 27:23:16.86 \\
4604.01 &                 KP &                     1.56 &           2.23 &                          476.88 &                    0.92 &                  90.06 & 05:05:47.03 & 21:32:53.52 \\
4610.01 &                 PC &                     1.56 &           3.11 &                          114.27 &                    0.69 &                  47.91 & 05:16:10.38 & 30:35:06.26 \\
5087.01 &                 KP &                     3.38 &          17.31 &                           10.89 &                    0.77 &                  59.25 & 04:29:39.09 & 22:52:57.24 \\
5129.01 &                 PC &                     3.64 &           7.41 &                           57.53 &                    1.19 &                 201.91 & 06:38:48.67 & 29:05:21.56 \\
\toprule
1459.01 &                 PC &                     2.49 &           9.16 &                           66.15 &                    0.82 &                 101.36 & 01:17:26.83 & 26:44:45.42 \\
1468.01 &                 CP &                     2.01 &          15.53 &                            2.14 &                    0.37 &                  24.74 & 01:06:36.93 & 19:13:29.71 \\
1471.01 &                 PC &                     3.92 &          20.77 &                           37.34 &                    0.97 &                  67.55 &  02:03:37.2 & 21:16:52.78 \\
4511.01 &                 PC &                     3.09 &          20.90 &                           42.72 &                    1.00 &                 121.83 & 03:17:13.27 & 15:30:06.22 \\
4524.01 &                 CP &                     1.69 &           0.93 &                          876.25 &                    1.11 &                  63.68 & 03:16:42.75 & 15:39:22.88 \\
4548.01 &                 PC &                     5.37 &           4.60 &                           84.87 &                    1.59 &                 165.59 & 02:25:21.87 & 25:31:50.44 \\
4607.01 &                 PC &                     3.08 &           5.51 &                          262.94 &                    1.31 &                 180.02 & 01:55:37.25 & 24:07:05.35 \\
4637.01 &                 PC &                     2.81 &          14.35 &                           38.49 &                    0.86 &                 112.15 & 02:13:03.56 &  19:24:09.6 \\
4639.01 &                 PC &                     2.88 &           3.99 &                          502.42 &                    1.03 &                 205.74 & 01:49:15.49 & 21:42:12.57 \\
4649.01 &                 PC &                     2.75 &          15.08 &                           68.57 &                    1.01 &                 148.25 & 01:59:49.57 &  16:20:48.1 \\
5076.01 &                 PC &                     3.13 &          23.44 &                           13.56 &                    0.85 &                  82.86 &  03:22:02.5 & 17:14:21.15 \\
5084.01 &                 PC &                     1.16 &           5.83 &                           37.73 &                    0.75 &                  21.36 & 03:03:49.09 & 20:06:38.12 \\
5319.01 &                 PC &                     3.75 &           4.08 &                           38.79 &                    0.48 &                  61.17 & 02:20:51.25 & 23:31:13.59 \\
5343.01 &                 PC &                     2.50 &          12.84 &                           39.96 &                    0.69 &                 120.86 & 03:12:06.25 & 24:32:00.82 \\
5358.01 &                 PC &                     2.92 &           2.66 &                          148.44 &                    0.80 &                 138.71 & 03:36:44.14 & 28:33:00.97 \\
5553.01 &                 PC &                     1.55 &           1.76 &                          439.12 &                    0.81 &                 103.35 & 02:52:00.52 & 15:03:20.39 \\
\toprule
\end{tabular}
\caption{Characteristics of primary sources observed in 2022--2023.  Columns as in Table \ref{tab-targets1}.}
\label{tab-targets2}
\end{table}

\clearpage
\FloatBarrier
\bibliography{seti20}

\end{document}